\documentclass[10pt,conference]{IEEEtran} 
\IEEEoverridecommandlockouts
\usepackage{url}
\def\BibTeX{{\rm B\kern-.05em{\sc i\kern-.025em b}\kern-.08em
		T\kern-.1667em\lower.7ex\hbox{E}\kern-.125emX}}

\usepackage{algorithm}
\usepackage[noend]{algpseudocode}
\usepackage{amsmath}
\usepackage{amsfonts}
\usepackage{amsthm}
\usepackage{balance}
\usepackage{graphicx}
\usepackage{xspace}
\usepackage{pifont}
\usepackage{multirow}
\usepackage{array}
\usepackage{booktabs}
\usepackage{microtype}
\usepackage{enumitem}
\usepackage{color,colortbl}
\usepackage{dblfloatfix}
\usepackage{framed}
\usepackage{subfigure}
\usepackage{xtab,afterpage}
\usepackage{tcolorbox}

\begingroup\expandafter\expandafter\expandafter\endgroup
\expandafter\ifx\csname IncludeInRelease\endcsname\relax
  \usepackage{fixltx2e}
\fi

\usepackage{listings}

\definecolor{darkgreen}{rgb}{0,0.42,0.24}
\lstdefinelanguage{diff}{
	morecomment=[f][\color{blue}]{@@},     
	morecomment=[f][\color{red}]-,         
	morecomment=[f][\color{darkgreen}]+,       
	morecomment=[f][\color{magenta}]{---}, 
	morecomment=[f][\color{magenta}]{+++},
}

\newcommand\lt[1]{{\lstinline+#1+}} 
\renewcommand\t[1]{{\lstinline+#1+}} 
\definecolor{dkgreen}{rgb}{0,0.5,0}
\definecolor{dkred}{rgb}{0.5,0,0}
\definecolor{gray}{rgb}{0.5,0.5,0.5}
\let\origthelstnumber\thelstnumber
\makeatletter
\newcommand*\Suppressnumber{%
  \lst@AddToHook{OnNewLine}{%
    \let\thelstnumber\relax%
     \advance\c@lstnumber-\@ne\relax%
    }%
}

\newcommand*\Reactivatenumber{%
  \lst@AddToHook{OnNewLine}{%
   \let\thelstnumber\origthelstnumber%
   \advance\c@lstnumber\@ne\relax}%
}

\definecolor{LightGray}{gray}{0.9}
\definecolor{Gray}{gray}{0.8}

\usepackage{tikz}

\newcommand{\sysname}{RAFL\xspace}
\newcommand{\sbir}{SBIR\xspace}
\newcommand{\irflname}{Blues\xspace}

\begin{document}
	
	\title{Better Automatic Program Repair\\by Using Bug Reports and Tests Together}
	
	\author{\IEEEauthorblockN{Manish Motwani and Yuriy Brun}
		\IEEEauthorblockA{
			University of Massachusetts \\
			Amherst, Massachusetts 01003-9264, USA \\
			\{mmotwani, brun\}@cs.umass.edu}
	}

	\maketitle
	
	\begin{abstract}
		
		Automated program repair is already deployed in industry, but concerns remain
		about repair quality. Recent research has shown that one of the main reasons
		repair tools produce incorrect (but seemingly correct) patches is imperfect
		fault localization~(FL). This paper demonstrates that combining information
		from natural-language bug reports and test executions when localizing faults
		can have a significant positive impact on repair quality. For example,
		existing repair tools with such FL are able to correctly repair 7~defects in
		the Defects4J benchmark that no prior tools have repaired correctly.
		
		We develop, \irflname, the first information-retrieval-based, statement-level FL technique
		that requires no training data. We further develop \sysname, the first
		unsupervised method for combining multiple FL techniques, which outperforms
		a supervised method. Using \sysname, we create \sbir by combining
		\irflname with a spectrum-based (SBFL) technique. Evaluated on 815~real-world
		defects, \sbir consistently ranks buggy statements higher than its 
		underlying techniques.
		
		We then modify three state-of-the-art repair tools, Arja, SequenceR, and
		SimFix, to use \sbir, SBFL, and \irflname as their internal FL. We evaluate
		the quality of the produced patches on 689~real-world defects. Arja and
		SequenceR significantly benefit from \sbir: Arja using \sbir correctly
		repairs 28~defects, but only 21 using SBFL, and only 15 using \irflname;
		SequenceR using \sbir correctly repairs 12~defects, but only 10 using SBFL,
		and only 4 using \irflname. SimFix, (which has internal mechanisms to
		overcome poor FL), correctly repairs 30~defects using \sbir and SBFL, but
		only 13 using \irflname. Our work is the first investigation of
		simultaneously using multiple software artifacts for automated program
		repair, and our promising findings suggest future research in this directions
		is likely to be fruitful.
		
	\end{abstract}
	
	\maketitle

	\pagestyle{empty}
	\thispagestyle{empty}
	
	\section{Introduction}
	\label{sec:Intro}
	
	Automated program repair~(APR) aims to reduce the cost of fixing bugs by
	automatically producing patches~\cite{Gazzola19, LeGoues19}. APR tools have
	been successful enough to be used in industry~\cite{Bader19, Marginean19,
		Noda20, Kirbas21}. Unfortunately, repair tools patch only a small fraction of
	defects correctly~\cite{Noda20, Motwani22, Smith15fse} and industrial
	deployments require significant manual oversight. Recent studies show that
	accuracy of the fault localization~(FL) used by APR has a significant effect
	on APR's success~\cite{Afzal21, Liu19, Jiang19, Wen17, Assiri17, Yang18}, and
	manually improving FL can correctly patch more defects~\cite{Afzal21, Liu21}.
	Some APR tools, such as SimFix~\cite{Jiang18} use tool-specific methods to
	address inaccurate FL; however, these methods are tightly coupled to the
	specific repair technique and not reusable by other tools. 
	
	Existing APR techniques use either developer-written test suites or
	natural-language bug reports. For the former, spectrum-based fault localization
	(SBFL) executes the tests and collects coverage information to identify suspicious
	statements. For the latter, information-retrieval-based fault localization~(IRFL)
	computes suspiciousness from the similarity between bug reports and program
	source. The defects these two types of APR tools repair tend to be
	complementary: For example, IRFL-based iFixR patches defects that
	16~SBFL-based repair tools cannot, and vice versa~\cite{Koyuncu19}. Further,
	combining multiple FL techniques can improve localization~\cite{Zou19, Li19,
		Le16issta}. We, therefore, hypothesize that combining SBFL and IRFL can
	improve APR. To test this hypothesis, we develop a novel IRFL technique and a
	novel method for combining FL techniques in an unsupervised fashion, and
	evaluate an SBFL, our IRFL, and the combined techniques in three
	state-of-the-art APR tools that have varied sensitivity to FL accuracy. 
	
	\begin{tcolorbox}
		Our main contribution is \sbir, a novel, reusable FL technique that combines
		bug reports and tests. The use of \sbir in APR is the first instance of APR
		simultaneously using multiple software artifacts, suggesting a promising new
		research direction.
		Our main finding is that the answer to the question ``Does FL that combines
		bug reports and tests improve APR performance?'' is a resounding yes, for
		many APR techniques. For example, on the latest Defects4J benchmark, 
		we correctly repair 7~defects
		that none of 14 prior APR tools could repair correctly~\cite{Liu19}.
	\end{tcolorbox}

	\smallskip
	\noindent
	\textbf{Contributions on using bug-reports.} We create \irflname,
	(Section~\ref{sec:irflapproach}), the first reusable, APR-agnostic,
	unsupervised, statement-level IRFL technique that localizes defects using bug
	reports. 
	Prior IRFL techniques are either file- or
	method-level~\cite{Zhou12, Saha13, Wong14, Youm15, Wen16, Koyuncu19arxiv}, or
	is the technique used internally by iFixR~\cite{Koyuncu19}. iFixR's FL
	requires hard-to-get training data and is tightly coupled to its APR
	implementation~\cite{Koyuncu19}. Unlike iFixR's FL, \irflname can localize
	defects to all 57~kinds of Java AST~expressions, (iFixR only handles
	5~\cite{Koyuncu19}). We empirically demonstrate that \irflname
	outperforms iFixR's FL (Section~\ref{sec:blues_evaluation}).

	\smallskip
	\noindent
	\textbf{Using tests.}
	Our SBFL technique is not novel. We implement SBFL using the latest version~(v1.7.2) of
	GZoltar, and the Ochiai ranking strategy, which is one of the most effective
	ranking strategies in object-oriented programs~\cite{Xuan14icsme, Zou19}, and
	is used by most test-suite-based repair tools~\cite{Liu19}
	(Section~\ref{sec:sbflapproach}).
	
	\smallskip
	\noindent
	\textbf{Contributions on using bug reports and tests together.}
	To combine FL techniques, we develop \sysname,
	(Section~\ref{sec:combinedfl}), a novel approach inspired by search-based
	software engineering~\cite{Harman01} that uses rank aggregation
	algorithms~\cite{Lin10} to combine multiple ranked lists of top-k suspicious
	statements obtained by different FL techniques. While \sysname can combine
	any FL~techniques, we focus on combining SBFL and IRFL, which are used
	separately by existing repair tools. We use \sysname to develop \sbir that
	uses the cross-entropy Monte Carlo algorithm~\cite{Rubinstein13} and the
	Spearman Footrule distance~\cite{Brandenburg13} to combine our SBFL and
	\irflname. We evaluate our SBFL, \irflname, and \sbir on
	815~real-world defects in the Defects4J~(v2.0)~\cite{Gay20} benchmark (out of
	the benchmark's 835~defects, 18 have no bug reports, and 2 have irrelevant
	test execution information) and find that \sbir consistently outperforms the
	underlying techniques (Section~\ref{sec:sbir_results}). While one could use
	existing \emph{supervised} combining FL techniques (e.g.,
	CombineFL~\cite{Zou19}, DeepFL~\cite{Li19}, Fluccs~\cite{Sohn17},
	Savant~\cite{Le16issta}, Multric~\cite{Xuan14icsme}, and TraPT~\cite{Li17}),
	our study elects to use a new, \emph{unsupervised} method because the prior
	techniques were trained on Defects4J and thus cannot be applied to an
	evaluation on Defects4J. Retraining the supervised techniques poses complex
	technical challenges, requires a large, independent, annotated dataset that
	simply doesn't exist today, and does not guarantee previously observed
	performance. We demonstrate that our unsupervised technique outperforms
	existing supervised ones (Section~\ref{sec:sbir_vs_soa}).
	
	\begin{tcolorbox}
		Importantly, existing supervised methods for both IRFL and for combining
		multiple FL techniques require extensive training data, which is expensive
		(sometimes prohibitively so) to obtain. Our evaluation shows that our
		\emph{unsupervised} methods (\irflname and \sysname) consistently perform as
		well as or better than the supervised methods, without needing the expensive
		training data.
	\end{tcolorbox}
	
	\smallskip
	\noindent
	\textbf{Contributions on effect on APR.}
	To study the effect of combining FL on repair quality (Section~\ref{sec:APR
		Evaluation Tools, Dataset, and Metrics}), we select Arja~\cite{Yuan20},
	SequenceR~\cite{Chen19}, and SimFix~\cite{Jiang18}, three state-of-the-art
	APR tools that have varied FL~sensitivity~\cite{Liu21},
	are applicable to general defects, use varied repair
	approaches, and have public implementations.
	We evaluate these tools using our SBFL, \irflname, and \sbir
	FL techniques on the 689~single-file-edit defects in the Defects4J~(v2.0)
	benchmark, and find that \sbir enables APR to repair more defects correctly.
	For tools that have been shown to be more sensitive to FL~\cite{Liu21}, \sbir
	significantly improves patch quality (Section~\ref{sec:apr_results}).
	
	Our evaluation answers four research questions:
	
	\begin{tcolorbox}
		
		\textbf{RQ1. Does \irflname localize defects better than existing
			approaches?} Yes. \irflname consistently ranks buggy statements higher than
		state-of-the-art iFixR's supervised IRFL technique
		(Section~\ref{sec:blues_evaluation}).
		
	\end{tcolorbox}
	
	\begin{tcolorbox}

		\textbf{RQ2. Does \sbir improve FL over the techniques it is composed of?}
		Yes. \sbir consistently ranks buggy statements 
		higher than its underlying SBFL and \irflname (Section~\ref{sec:sbir_results}).
		
		\textbf{RQ3. Does \sbir outperform state-of-the-art FL?} Yes. \sbir
		consistently ranks buggy statements higher than 9~standalone FL techniques and an
		existing supervised FL-combining method 
		(Section~\ref{sec:sbir_vs_soa}).
		
		\textbf{RQ4. Does \sbir improve repair quality?} Yes. \sbir enables repairing
		more defects correctly for Arja and SequenceR (the more FL-sensitive tools).
		For example, Arja using \sbir correctly repairs 28~defects, but only 21 using
		SBFL, and only 15 using \irflname. In fact, using \sbir, Arja repairs
		7~defects it cannot repair with either SBFL or \irflname, suggesting that the
		combination of bug reports and tests is even more useful, at times, than
		using both types of information separately. SimFix already has internal
		mechanisms for dealing with poor FL, and correctly repairs 30 defects using
		both \sbir and SBFL, but only 13 using \irflname. We empirically show that \sbir
		significantly reduces repair failures due to localization errors. Finally,
		using \sbir, these tools correctly repair 7 defects that none of prior 14 APR
		tools repaired correctly, representing a 7.5\% improvement in the number
		of defects ever repaired correctly by APR (Section~\ref{sec:apr_results}).
	\end{tcolorbox}
	
	APR has already shown effectiveness in real-world
	scenarios, but producing correct repairs is one of the remaining hurdles
	preventing wide deployment in industry~\cite{Gazzola19}. This paper makes
	progress addressing this challenge by (1)~developing a new FL
	technique suitable for APR that uses both bug reports and tests,
	demonstrating that it localizes defects better than techniques 
	that use only bug reports or only tests, and (2)~demonstrating
	that with this new FL, APR tools can repair more defects correctly. 
	
	We make all of our data, source code, and documentation to reproduce 
	our results publicly available~\cite{MotwaniSBIRArtifact}.
	The rest of this paper is organized as follows.
	Section~\ref{sec:approach} describes our FL techniques and
	Section~\ref{sec:evaluation} evaluates the FL techniques, and their effect
	on APR.
	Section~\ref{sec:relatedwork} places our work in the context of related research, and
	Section~\ref{sec:contributions} summarizes our contributions.
	
	\section{Combining FL for Program Repair}
	\label{sec:approach}
	
	This section describes our \irflname and SBFL techniques, our method for
	combining FL techniques called \sysname, and using \sysname to combine
	\irflname and SBFL into \sbir.
	
	\subsection{\irflname: Localizing Bugs Using Bug Reports}
	\label{sec:irflapproach}
	
	We design \irflname, an IRFL technique that uses bug reports to localize
	defects at the statement level. We create our own technique because existing
	techniques~\cite{Zhou12, Saha13, Wong14, Youm15, Wen16, Koyuncu19arxiv}
	localize defects at the file or method level, while APR tools require
	statement-level localization. We do not use iFixR's~\cite{Koyuncu19} IRFL
	(the only existing statement-level IRFL) because its pre-trained model uses
	projects~\cite{Lee18} that overlap with the Defects4J and retraining on
	independent projects poses complex technical challenges and requires
	\emph{another} large annotated dataset of real-world defects. Further, iFixR
	ignores \texttt{for} and \texttt{while} loops, which \irflname handles.
	\irflname builds on BLUiR~\cite{Saha13}, an unsupervised file-level IRFL
	technique that uses structured information retrieval to compute the
	similarity between bug reports and source code files. We select BLUiR because
	it is efficient, does not require training data, and performs comparably
	to other state-of-the-art file-level IRFL techniques~\cite{Lee18}.
	Algorithm~\ref{alg:blues} describes our \irflname approach.
	
	\begin{algorithm}[b]
		\scriptsize
		\caption{Blues: Statement-level IR-based FL.}
		\hspace*{\algorithmicindent}\textbf{Input:} $\textit{br}$: a bug report URL \\
		\hspace*{\algorithmicindent}\textbf{Input:} $\textit{srcFiles}$: collection of source files  \\
		\hspace*{\algorithmicindent}\textbf{Input:} $\textit{irTool}$: file-level IRFL tool \\
		\hspace*{\algorithmicindent}\textbf{Input:} $f$: number of suspicious files to consider \\
		\hspace*{\algorithmicindent}\textbf{Input:} $m$: number of suspicious statements per file to consider \\
		\hspace*{\algorithmicindent}\textbf{Input:} $\textit{ScoreFn}$: function to combine file and statement scores \\
		\hspace*{\algorithmicindent}\textbf{Output:} $\textit{rankedStmtList}$: ranked list of suspicious statements \\
		\begin{minipage}{\columnwidth}
			\begin{algorithmic}	[1]
				\Function {Main } {$\textit{br}$, $\textit{srcFiles}$, $\textit{irTool}$, $f$, $m$, $\textit{ScoreFn}$}
				
				\State $\textit{br\_xml} \gets$ ParseBugReportAndConvertToXML($\textit{br}$) 
			
				\State $\textit{src\_files\_xml} \gets$ ParseSrcFilesAndConvertToXML($\textit{srcFiles}$)
			
				\State PreProcess($\textit{br\_xml}$)
			
				\State PreProcess($\textit{src\_files\_xml}$)
				\State $\textit{ranked\_files} \gets$ Okapi($\textit{br\_xml}$, $\textit{src\_files\_xml}$, $\textit{irTool}$)
			
				\State $\textit{ranked\_stmts} \gets$ LocalizeStatements($\textit{br\_xml}$, $\textit{ranked\_files}$, $\textit{irTool}$)
				\State $\textit{rankedStmtList} \gets$ Ranker($\textit{ranked\_files}$, $\textit{ranked\_stmts}$, $f$, $m$, $\textit{ScoreFn}$) 
			
				\\
				\hspace{2.5ex}	\Return $\textit{rankedStmtList}$
				\EndFunction
				\\
				\Function {LocalizeStatements }{$\textit{br\_xml}$, $\textit{ranked\_files}$, $\textit{irTool}$}
				\State $\textit{src\_stmts\_xml} \gets [~]$ \Comment{stores XMLs of parsed source statements}
				\For {$f \in \textit{ranked\_files}$}
				\State $S \gets$ extractASTStatements($f$) \Comment{extract 57~kinds of Java AST statements}
				\For {$\textit{ast\_stmt} \in S$}
				\State $\textit{stmt\_xml} \gets$ ParseStmtAndConvertToXML($\textit{ast\_stmt}$, $f$) 
			
				\State $\textit{src\_stmts\_xml.append(stmt\_xml)}$
				\EndFor
				\EndFor 
				
				\State PreProcess($\textit{src\_stmts\_xml}$)
				
				\State $\textit{ranked\_stmts} \gets$ Okapi($\textit{br\_xml}$, $\textit{src\_stmts\_xml}$, $\textit{irTool}$)
			
				\\
				\hspace{2.5ex}	\Return $\textit{ranked\_stmts}$
				\EndFunction
				
			\end{algorithmic}
		\end{minipage}
		\label{alg:blues}
	\end{algorithm}

	\textbf{Ranking Suspicious Files.} 
	For each defect, \irflname' inputs are the bug report URL, the source files, 
	the number of top ranked files to consider, the number of top ranked 
	statements per file to consider, and a function to combine statement 
	and file suspiciousness scores.
	\irflname crawls the bug report from the input URL and
	parses the bug report to extract identifiers from the
	\emph{summary} and \emph{description} fields, and stores the information in
	a separate structured XML document (line\,2 in Algorithm~\ref{alg:blues}). 
	Next, \irflname processes the abstract syntax tree~(AST) of source files 
	to extract identifiers associated with comments and with class, method, and
	variable names, and stores them in separate XML documents (line\,3).
	\irflname preprocesses the terms stored in all the XML documents 
	using CamelCase splitting, which improves matching recall, 
	text normalization~(removes punctuation, performs case-folding, tokenizes terms), 
	stopword removal~(removes extraneous terms),
	and stemming~(conflates variants of the same underlying term)
	(lines\,4--5).
	\irflname then feeds the bug report and source file XML documents to BLUiR 
	to compute ranked lists of suspicious files (line\,6).
	BLUiR uses an IR model (TF-IDF formulation based on
	the BM25 (Okapi) model~\cite{Robertson2000}) to search and rank the
	files based on their similarity with the bug report.
	\irflname uses the same tuning parameters as BLUiR, which prior
	work~\cite{Saha13} tuned using AspectsJ that does not overlap with Defects4J.
	
	\textbf{Ranking suspicious statements.}
	To rank suspicious statements from the top-ranked suspicious files,
	\irflname parses the ASTs of the top-ranked 
	suspicious files to extract 57~types of Java AST statements 
	(lines\,13--14). 
	Prior work~\cite{Liu18icsme} shows that localizing bugs at the expression-level
	can improve repair tools. Therefore, unlike iFixR~\cite{Koyuncu19}, 
	which only extracts five kinds of AST~statements (If, Return, Expression,
	FieldDeclaration, and VariableDeclaration), \irflname extracts
	32~AST~expressions~\cite{ASTExpr},  
	3~AST~nodes (SingleVariableDeclaration, 
	AnonymousClassDeclaration, Annotation), and 22~AST~statements~\cite{ASTStmt},  
	17 of which iFixR ignores, including \texttt{for} loops, \texttt{while}
	loops, \texttt{do} statements, etc. For readability, we refer to the
	AST~expression, AST~node, and AST~statement as \emph{statement}.
	
	For each statement, \irflname identifies its line number in the associated
	source file along with the file name, extracts identifier terms, and stores
	this information in an XML document (lines\,15--16). \irflname creates these
	XML documents for all the statements extracted from the ranked source files (line\,17),
	and preprocesses these XMLs (line\,18) in the same way it pre-processes
	source file XMLs. Next, \irflname feeds these statement and the bug report
	XMLs to BLUiR that outputs a ranked list of the statements. \irflname
	extracts the line number, source file name, and suspiciousness scores from
	the output to create a ranked list of suspicious statements (line\,19). Note
	that these ranked statements do not consider the ranks of their associated
	source files. Real-world projects contain many source files, and our
	experiments show that treating all statements in a higher-ranked file to be
	more suspicious than the ones in lower-ranked files is sometimes suboptimal,
	so we also explore other strategies. To combine the ranked suspicious files
	and statements, \irflname provides a ranker module that uses the three
	parameters: $f$, the number of suspicious files to consider; $m$, the number
	of suspicious statements per file to consider; and $\mathit{ScoreFn}$, a
	function for combining the file and statement suspiciousness scores
	(line\,8). We define two such functions: $\mathit{Score}_{\mathit{high}}$
	ranks the $m$ most suspicious statements in the most suspicious file,
	followed by $m$ statements in the next file, and so forth.
	$\mathit{Score}_{\mathit{wt}}$ uses the files' scores as weights for the
	associated suspicious statements and recomputes the weighted suspiciousness
	scores by multiplying the scores of the statements with the score of the
	associated file. We set $f = 50$ based on the recommendation of a prior
	study~\cite{Koyuncu19}. We run \irflname' ranker module using six different
	configurations: five ($m \in \{1, 25, 50, 100, \mathit{all}\}$) with
	$\mathit{Score}_{\mathit{high}}$, and one ($m = \mathit{all}$) with
	$\mathit{Score}_{\mathit{wt}}$. For each of the six configurations, \irflname
	produces a ranked list of statements.
	 
	We found that the six configurations localize complementary defects, so we
	use Algorithm~\ref{alg:blues_ensemble} to combine the six ranked lists into a
	single list, which we call \emph{\irflname ensemble}. The algorithm to
	combine lists sorts the statements using each statement's highest rank in the
	six lists, breaking ties using the number of lists in which the statement
	occurs (line\,15 in Algorithm~\ref{alg:blues_ensemble}). To fairly compare
	suspiciousness scores across lists, the algorithm normalizes the scores first
	(line\,5).
	Note that computing the individual configurations and the ensemble is a
	relatively low-cost process. One only needs to rerun \irflname's ranker
	module (line\,8 in Algorithm~\ref{alg:blues}) and
	Algorithm~\ref{alg:blues_ensemble}, not the entire \irflname pipeline. From
	here on, we use only the ensemble and refer to it as just \irflname.
	
	\begin{algorithm}[t]
		\caption{Combining ranked suspicious statement lists using suspiciousness
			scores and consensus.}
		\footnotesize
		\hspace*{\algorithmicindent}\textbf{Input:} $\textit{rankedStmtLists} \gets [l_1, l_2, \ldots]$  \\
		\hspace*{\algorithmicindent}\textbf{Output:}  $\textit{combinedStmtList}$ \\
		\begin{minipage}{\columnwidth}
			\begin{algorithmic}[1]
				\Function {CombineLists}{$\textit{rankedStmtLists}$}
				\State $\textit{stmt\_maxscore} \gets \{\}$	\Comment{max susp. score of stmt from all the lists}
				\State $\textit{stmt\_listcount} \gets \{\}$ \Comment{number of lists in which a stmt occurs}
				\For {$l_{k} \in rankedStmtLists$} 
				\State $\textit{list}_{n} \gets$ NormalizeScoresInList($l_{k}$) 
				\For {$(\textit{stmt}, \textit{score}) \in \textit{list}_{n}$}	
				\If {$\textit{score} > 0.0$}
				\If {$\textit{stmt} \notin \textit{stmt\_listcount}$}	
				\State $\textit{stmt\_listcount[stmt]} = 1$
				\State $\textit{stmt\_maxscore[stmt]} = \textit{score}$
				\Else	\Comment{stmt seen before, update maxscore if needed}					
				\State $\textit{stmt\_listcount[stmt]} \mathrel{+}= 1$ 
				\If {$\textit{score} > \textit{stmt\_maxscore[stmt]}$}
				\State	$\textit{stmt\_maxscore[stmt]} = \textit{score}$	
				\EndIf
				\EndIf
				\EndIf
				\EndFor		
				\EndFor
				\State $\textit{combinedStmtList} \gets$ \Call{sort}{$\textit{stmt\_maxscore}$, $\textit{stmt\_count}$} 
				\\
				\hspace{3ex}\Return $\textit{combinedStmtList}$
				\EndFunction
			\end{algorithmic}
		\end{minipage}
		\label{alg:blues_ensemble}
	\end{algorithm}

	\subsection{Spectrum-Based Fault Localization}
	\label{sec:sbflapproach}
	
	\looseness-1
	We do not create a new SBFL technique, but combine existing
	tools to produce a state-of-the-art implementation. SBFL compares program
	spectra\,---\,measurements of the runtime behavior of a program, such as code
	covered by tests~\cite{Harrold2000}\,---\,of passing and failing
	developer-written tests to rank program elements, such as 
	statements. SBFL calculates suspiciousness scores using a ranking
	strategy that considers four values collected from the spectrum:
	the number of failing tests that do~($e_f$) and do not~($n_f$) execute
	the element, and the number of passing
	tests that do~($e_p$) and do not~($n_p$) execute the element. While there are multiple ranking
	strategies, including Ochiai~\cite{Abreu07},
	DStar~\cite{Wong13}, and Tarantula~\cite{Jones05}, empirical
	studies~\cite{Xuan14icsme, Zou19} have found that Ochiai is more effective
	for object-oriented programs. Most SBFL-based APR tools use Ochiai, and
	so does our study.
	
	There exist multiple frameworks that APR tools use to compute code coverage,
	including JaCoCo~\cite{Hoffmann09}, GZoltar~\cite{Campos12}, and
	Cobertura~\cite{Christou15}. Our study uses GZoltar because 
	most APR tools use it, and a recent study comparing 14~APR tools 
	used multiple GZoltar versions, showing that the
	latest-at-the-time version (v1.6.0) significantly improved FL results and
	repair performance~\cite{Liu19}. We use the latest
	version~(v1.7.2) of GZoltar available at the time of running our experiments. GZoltar's inputs are the
	source code and test suite and its outputs are each statement's
	$e_f$, $n_f$, $e_p$, and $n_p$. We use the Ochiai ranking formula 
	to compute suspiciousness scores:
		$\mathit{score} = \frac{e_f}{\sqrt{(e_f + n_f)(e_f + e_p)}}$.
	
	To validate our SBFL implementation, we compare it to
	previously reported results~\cite{Liu19} on Defects4J (v1.2.0) for SBFL
	implemented using Ochiai and older versions of Gzoltar. 
	Figure~\ref{fig:sbfl_flresults} shows our SBFL implementation localizes 
	13 more defects than the best prior version.
	
	\begin{figure}[t]
		\begin{center}
			\small
			\resizebox{\columnwidth}{!}{%
				\begin{tabular}{lcccccc|c}
					\toprule
					project & Chart & Closure & Lang & Math & Mockito & Time & Total\\
					\#defects & 26 & 133  & 65  & 106 & 38 & 27 & 395 \\
					\midrule
					GZ~v0.1.1 & 22 & \phantom{0}78 & 29 & \phantom{0}91 & 21 & 22 & 263 \\ 
					GZ~v1.6.0 & 24 & \phantom{0}95 & 57 & 100 & 23 & 22 & 321 \\
					\textbf{GZ~v1.7.2} & $\mathbf{25}$ & $\mathbf{101}$ & $\mathbf{53}$ & \phantom{0}$\mathbf{96}$ & $\mathbf{36}$ & $\mathbf{23}$ & $\mathbf{334}$ \\
					\bottomrule
				\end{tabular}
			}
			\caption{Our SBFL (implemented using GZoltar~(v1.7.2) and Ochiai), in \textbf{bold}, localizes more defects than prior SBFLs using older versions of Gzoltar~\cite{Liu19}.}
			\label{fig:sbfl_flresults}
		\end{center}
	\end{figure}
	
	In the remainder of this paper, when we refer to our SBFL, we are referring to
	this particular implementation.
	
	\subsection{Combining FL Techniques}
	\label{sec:combinedfl}
	
	Existing approaches to combining multiple FL techniques~\cite{Li19, Zou19,
		Sohn17, Le16issta, Xuan14icsme} typically use \emph{learning to
		rank}~\cite{Burges05} supervised machine learning. These techniques use
	multiple FL techniques' suspiciousness scores as \emph{features} to train a
	model to rank buggy statements higher than non-buggy ones. Such approaches
	require a training dataset of program statements annotated with
	suspiciousness scores from multiple FL techniques, and the manually
	labeled ground truth ``buggy'' or ``not-buggy''. 
	Such training data is hard to create because of the required manual effort,
	and the performance of trained models depends heavily on its data and
	features~\cite{Mahalakshmi15}.
	
	Instead, we propose \sysname, a novel unsupervised approach that requires no
	training. We formulate the problem of combining different FL techniques as a
	rank aggregation~(RA)~\cite{Lin10} problem. RA involves combining multiple
	ranked lists (base rankers) into one ranked list (aggregated
	ranker)~\cite{Deng14}. The RA problem has been studied extensively in
	information retrieval~\cite{Dwork01www}, marketing and advertisement
	research~\cite{Lin10}, social choice (elections)~\cite{Dwork01www}, and
	genomics~\cite{Kolde12}. We propose to use RA algorithms to combine multiple
	FL techniques' ranked lists. We next describe our \sysname approach to
	combine FL techniques~(Section~\ref{sec:rafl}) and using it to combine
	\irflname and SBFL~(Section~\ref{sec:sbir}). Section~\ref{sec:sbir_vs_soa}
	will empirically show that our approach outperforms the supervised ones.
	
	\subsubsection{\sysname: Rank Aggregation-based FL}
	\label{sec:rafl}
	
	\looseness-1
	FL techniques typically assign suspiciousness scores to hundreds of program
	statements. Combining multiple ranked lists, which are often inconsistent,
	such that the result is as close as possible to the individual lists
	according to some distance metric, can become combinatorially intractable. We
	propose rank aggregation-based FL~(\sysname), a novel approach that uses RA
	algorithms to combine FL. Our technique takes inspiration from the research
	in search-based software engineering~\cite{Harman01}, which involves applying
	metaheuristic search techniques to solve problems of balancing competing (and
	sometimes inconsistent) constraints. \sysname works as follows. Let $L_1,
	L_2, \ldots, L_m$ be $m$ ordered lists of suspicious statements (e.g.,
	obtained using $m$ FL techniques). \sysname aims to create an ordered
	list~$\delta$ of length~$k\geq 1$ that combines the statements in the
	individual lists by minimizing the weighted sum of the distances between
	$\delta$ and the individual lists. Formally, \sysname minimizes the objective
	function $f(\delta) = \sum_{i=1}^{m} w_{i}d(\delta, L_{i})$, where $w_{i}$ is
	the importance weight associated with list $L_{i}$, and $d$ is a distance
	metric.
	
	To minimize the objective function, \sysname samples multiple lists of $k$~statements from
	the unique statements in the individual lists, using an algorithm-specific 
	sampling strategy. 
	\sysname computes the objective function for each sampled list. Iteratively,
	\sysname updates the sampled lists using the objective function computations,
	e.g., by adjusting the sampling probabilities or using genetic algorithms to
	select the next generation of sampled lists.
	This iteration continues until \sysname observes no change in the objective
	function scores for a fixed number of iterations, returning the
	lowest-scoring list.
	
	\looseness-1
	Our \sysname implementation uses the RankAggreg~\cite{Pihur09} package, which
	implements several RA algorithms (cross-entropy Monte Carlo~(CE),
	genetic algorithm~(GA), and brute force) and provides distance
	metrics (Spearman Footrule~\cite{Brandenburg13}, and Kendall's tau~\cite{Brandenburg13Kendall}).
	The left two columns in Figure~\ref{fig:rafl_params} list \sysname
	configuration parameters, which can be used to select combinations of RA
	algorithms and distance metrics to combine FL.  
	
	\begin{figure}[t]
		\begin{center}
			\footnotesize
			\begin{tabular}{lp{0.5\linewidth}p{0.2\linewidth}} 
				\toprule
				parameter & definition & \sbir value \\ 	
				\midrule
				k 		  & size of the combined list & 100 \\
				seed      & seed specified for reproducibility & 1 \\
				distance  & Spearman or Kendall & Spearman \\ 
				method	  & algorithm~(CE or GA) & CE \\
				maxIter   &	max \#iterations allowed (default 1000) & 1000 \\
				convIn 	  & \#consecutive iterations to decide if algorithm has converged (default: 7 for CE, 30 for GA) & 7 \\
				importance & vector of weights ($w_{i}$) indicating the importance of each list (default: a vector of
				1's (equal weights to all lists)) & default \\
				N		  & \#samples generated in each iteration. Used only by the CE (default: $10kn$, where $n$ is the \#unique statements considering all ranked lists and $n~>>~k$, otherwise at least $k^2$) & 10,000 \\
				$\rho$		& ($\rho \cdot N$) is quantile of candidate lists sorted by the objective function scores. 
				Used only by the CE. (default: 0.01 when $N \geq 100$ and 0.1 otherwise) &  0.01\\
				popSize	& population size in each generation for the GA (default 100) & NA \\
				CP		& Cross-over probability for the GA (default 0.4)	& NA \\
				MP		& Mutation probability for the GA	& NA \\
				\bottomrule	
			\end{tabular}
			\caption{\sysname configuration parameters.}
			\label{fig:rafl_params}
		\end{center}
	\end{figure}

	\subsubsection{\sbir: Combining  \irflname and SBFL}
	\label{sec:sbir}
	
	To combine the suspicious statement lists from
	\irflname~(Section~\ref{sec:irflapproach}) and our
	SBFL~(Section~\ref{sec:sbflapproach}), we use \sysname to develop \sbir using
	the cross-entropy Monte Carlo~(CE) rank aggregation algorithm with the
	Spearman Footrule distance. We make these choices because prior work found CE
	to be typically more efficient than genetic algorithms~\cite{Pihur20} and
	than Borda count~\cite{Debroy13, Pihur09}, and because computing the Spearman
	Footrule distance is faster than Kendall's tau.
	
	The CE algorithm represents an ordered list of $k$~statements
	using a 0--1 matrix of size $n \times k$, where $n$ is the total number 
	of unique statements in the ranked lists and $k$ is the length of the desired 
	combined list. The algorithm imposes two constraints: each column 
	sums up to exactly 1, and each row sums up to at most 1. 
	Under this representation, an ordered list of size $k$ is uniquely determined 
	by reordering the matrix' rows (statements) such that the top $k$ rows form
	the identity matrix.
	For example, if the full list was [A, B, C], a $3 \times 2$ matrix,
	{\tiny
		$
		\begin{bmatrix}
			0 & 0  \\
			0 & 1  \\
			1 & 0  \\
		\end{bmatrix}
		$}
	would translate into the candidate top 2 list of (C, B).
	
	\looseness-1
	CE algorithm's goal is to identify a matrix that results 
	in the minimum objective function score out of all possible matrices.
	The CE algorithm uses the following four steps:
	(1)~\textbf{Initialization} creates an $n \times k$ matrix 
	and assigns each cell a probability of $\frac{1}{n}$.
	This matrix represents the multinomial sampling probabilities of the statements: each
	statement (row) is equally likely to be in each of the $k$ positions (column).
	Next, CE runs steps~2~and~3 iteratively.
	(2)~\textbf{Sampling} generates $N$ 0--1 matrices using the restricted~(truncated) 
	multinomial sampling~\cite{Sanathanan77} using the current probabilities. 
	The output of this step are $N$ (new) randomly generated 0--1 matrices of 
	size $n \times k$.
	(3)~\textbf{Updating} computes the objective function scores for each of the
	$N$ sampled matrices, sorts the sampled matrices in the ascending order of
	the scores, and identifies $\rho$-quantiles $y^t$ of the sorted matrices.
	The algorithm uses the objective function scores of the matrices in iteration~$t$
	to update the multinomial cell probabilities of unique statements 
	that tend to minimize the objective function scores of the matrices
	sampled in the next iteration, as follows: 
	\begin{equation*}
		p_{jr}^{t+1} = (1-w)p_{jr}^{t} + w \frac{\sum_{i=1}^{N} \mathit{I(f(\delta_{i}) \leq y^t)x_{ijr}}}{\sum_{i=1}^{N} \mathit{I(f(\delta_{i}) \leq y^t)}}
	\end{equation*}
	
	\noindent where $1 \leq j \leq n$, $1 \leq r \leq k$,
	$p_{jr}^{t}$ is the probability of the unique statement at the $jr^{\text{th}}$ position in the matrix
	at iteration $t$ and $p_{jr}^{t+1}$ is its updated value at iteration $t+1$; 
	$f(\delta_{i})$ is the objective function score of the $i^{\text{th}}$
	sampled matrix and $x_{ijr}$ is the value of the $jr^{\text{th}}$ cell
	of the $i^{\text{th}}$ sampled matrix;
	$w$ is a weight parameter with a default value of 0.25 (tuned by prior work~\cite{Pihur20} on independent dataset) and $I$ is the indicator function. 
	(4)~\textbf{Convergence} stops the iteration when the minimum value of the objective 
	function does not change in a preset number of iterations.
	The matrix with a minimum objective function score in the final iteration represents 
	the final combined list of statements.
	
	\looseness-1
	\sbir combines SBFL's and \irflname' ranked suspicious statement
	lists to produce a single list of top-100 statements.
	The right column in Figure~\ref{fig:rafl_params} shows the values of configuration 
	parameters we used to develop \sbir. We select $k = 100$ because most APR tools consider 
	at most 100~statements during repair.
	We set $w_{i} = 1.0$ to assign equal importance to SBFL and \irflname
	and use default values of other parameters including 
	$w$ (used in updating sampling probabilities), and $\rho$ that 
	are tuned by prior work~\cite{Pihur20} on a dataset that does not overlap with Defects4J.
	
	\section{Evaluation}
	\label{sec:evaluation}
	
	We next evaluate our FL techniques and their effect on APR.
	
	\begin{figure}[t]
		\begin{center}
			\resizebox{\columnwidth}{!}{%
				\begin{tabular}{@{}ll>{\raggedright\arraybackslash}p{0.45\columnwidth}rrr@{}}
					\toprule 
					\textbf{identifier} & \textbf{project}  & \textbf{description} & $\!\!\!\!\!\!\!\!\!\!\!\!\!$\textbf{all} & \textbf{sfd} & \textbf{sld}\\
					\midrule
					Chart           & jfreechart            & framework to create charts & \phantom{00}8   & \phantom{00}8 & \phantom{00}4 \\
					Cli             & commons-cli           & API for parsing command line options & \phantom{0}39   & \phantom{0}32 & \phantom{00}3 \\ 
					Closure         & closure-compiler      & JavaScript compiler & 174  & 137  & \phantom{0}23 \\   
					Codec           & commons-codec         & implementations of encoders \& decoders  & \phantom{0}18  & \phantom{0}14 & \phantom{00}8  \\
					Collections     & commons-collections   & Java Collections Framework extensions & \phantom{00}4  & \phantom{00}4 & \phantom{00}1    \\   
					Compress        & commons-compress      & API for file compression utilities & \phantom{0}47  & \phantom{0}43 & \phantom{00}4   \\   
					Csv             & commons-csv           & API to read and write CSV files & \phantom{0}16  & \phantom{0}15 & \phantom{00}5   \\
					Gson            & gson                  & API to convert Java Objects into JSON & \phantom{0}18  & \phantom{0}16 & \phantom{00}2   \\
					JacksonCore     & jackson-core          & core part of the Java JSON API (Jackson) & \phantom{0}26  & \phantom{0}19 & \phantom{00}3   \\
					JacksonDatabind & jackson-databind      & data-binding package for Jackson & 111  & \phantom{0}91 & \phantom{0}13 \\ 
					JacksonXml      & jackson-dataformat-xml & data format extension for Jackson & \phantom{00}6   & \phantom{00}6  & \phantom{00}1    \\
					Jsoup           & jsoup                  & HTML parser & \phantom{0}93  & \phantom{0}75  & \phantom{0}18   \\
					JxPath          & commons-jxpath         & XPath (an expression language) interpreter  & \phantom{0}22  & \phantom{0}13 & \phantom{00}1   \\
					Lang            & commons-lang           & extensions to Java Lang & \phantom{0}64  & \phantom{0}64 & \phantom{0}10   \\
					Math            & commons-math           & library of math utilities & 106   & \phantom{0}98 & \phantom{0}23 \\
					Mockito         & mockito                & a unit-test mocking framework & \phantom{0}38   & \phantom{0}33 & \phantom{00}7 \\
					Time            & joda-time              & date and time library & \phantom{0}25 & \phantom{0}21 & \phantom{00}3  \\
					\midrule
					total           &                        & & 815 & 689 & 129   \\
					\bottomrule
				\end{tabular}
			}
			
			\caption{The ``all'' column shows the 815~defects from the 17~real-world Java projects
				in the Defects4J~(v2.0) benchmark we use to evaluate our FL techniques. 
				The ``sfd'' column shows the 689~single-file-edit defects and the ``sld'' column shows 
				the 129~single-line-edit defects we use for APR evaluations.}
			\label{fig:d4j}
		\end{center}
	\end{figure}

	\subsection{FL Evaluation Dataset and Metrics}
	\label{sec:FL Evaluation Dataset and Metrics}
	
	We use the Defects4J~(v2.0)~\cite{Gay20} benchmark to evaluate our FL techniques.  Defects4J~(v2.0)
	targets Java~8 and consists of 835 reproducible defects from
	17~large open-source Java projects. Each defect comes with (1)~one buggy
	and one developer-repaired version of the project code with the changes
	minimized to those relevant to the defect; (2)~a set of developer-written
	tests, all of which pass on the developer-repaired version and at least one
	of which evidences the defect by failing on the buggy version; 
	and (3)~defect information, including the bug report URL. Out of the
	835~defects, 817 have the bug report URL available, making IRFL possible. For
	815 of the 817~defects, the test execution information was relevant 
	to make SBFL possible. 
	Figure~\ref{fig:d4j} describes these 815~defects, which we use to evaluate
	our FL techniques.
	
	\looseness-1
	We use two metrics, common to FL evaluations~\cite{Zou19}: (1)~$\mathit{hit}@k$
	is the number of defects localized in the top-k ranked
	statements, and (2)~EXAM is the fraction of ranked statements one has to 
	inspect before finding a buggy statement.
	$\mathit{hit}$@$k$ tells us how useful an FL technique is for APR 
	that uses the top~$k$ statements, while
	EXAM tells us how highly the buggy statements are ranked, 
	easing APR's job to produce correct patches. 
	
	Similar to prior studies~\cite{Liu19, Zou19, Koyuncu19}, we
	consider a defect successfully localized when at least one of the 
	buggy statements is in the top-k. 
	Unlike studies that break ties 
	by reassigning average rank~\cite{Pearson17} or expected rank~\cite{Zou19},
	we rank same-suspiciousness statements in the order they appear in
	the FL results, as this is how APR tools process them.
	
	\subsection{\irflname' Evaluation~\textbf{(RQ1)}}
	\label{sec:blues_evaluation}
	
	We next compare \irflname' performance to the state-of-the-art
	(Section~\ref{sec:blues_vs_ifixr}) and baseline
	(Section~\ref{sec:blues_vs_baseline}) IRFL techniques.
	
	\subsubsection{\irflname vs.\ State of the Art}
	\label{sec:blues_vs_ifixr}
	
	Figure~\ref{fig:bluesvsifixr} compares \irflname with iFixR's internal
	statement-level IRFL technique~\cite{Koyuncu19} on the 171~Lang
	and Math defects in Defects4J on which iFixR was
	evaluated\footnote{The iFixR FL results available at
		\url{https://github.com/TruX-DTF/iFixR/tree/master/data/stmtLoc} contain
		multiple statements with the same rank and multiple ranks for the same
		statement. We break ties by assigning the highest possible rank to each
		statement.}. As shown in Figure~\ref{fig:bluesvsifixr}, considering ranked
	lists of size $\geq 25$ (relevant for APR), \irflname consistently localizes
	more defects (higher $\mathit{hit}$@$k$) than iFixR's IRFL. Comparing the
	ranks of buggy statements in localized defects, \irflname places buggy
	statements higher (lowering EXAM) in the lists than iFixR. \irflname'
	advantage of using a lightweight unsupervised approach outweighs iFixR's
	supervised technique that requires 6~file-level IRFL techniques.
	
	\begin{figure}[t]
		\begin{center}
				\small
				\begin{tabular}{r|ccccc|c}
					\toprule
					(171 defects) & \multicolumn{5}{c}{$\mathit{hit}$@$k$} & EXAM \\
					\midrule
					& $k=$ 1  & 25 & 50  & 100 & all & $k=$ all \\
					\midrule
					iFixR     & $\mathbf{26}$ & $74$ & $95$ & $106$ & $135$ & $0.048$ \\
					\textbf{\irflname} & $11$ & $\mathbf{79}$ & $\mathbf{97}$ & $\mathbf{108}$ & $\mathbf{151}$ & $\mathbf{0.034}$ \\
					\bottomrule
				\end{tabular}
			\caption{For ranked lists of size $\geq 25$, 
				\irflname localizes more defects ($\mathit{hit}$@$k$) 
				and places buggy statements higher in the list (lower EXAM) 
				than the state-of-the-art IRFL technique used in iFixR 
				when evaluated on 171~Lang and Math defects in the Defects4J
				on which original iFixR was evaluated.}
			\label{fig:bluesvsifixr}
		\end{center}
	\end{figure}  
	
	\begin{figure}[t]
		\begin{center}
			\small
			\begin{tabular}{r|ccccc|c}
				\toprule
				(815 defects) & \multicolumn{5}{c|}{$\mathit{hit}$@$k$} & EXAM \\
				\midrule
				& $k=$ 1  & \phantom{0}25 & \phantom{0}50  & 100 & all & $k=$ all \\
				\midrule
				vanilla BLUiR & $ \phantom{0}26$ & $143$ & $192$ & $245$ & $611$ & 0.159 \\
				\textbf{\irflname} & $\phantom{0}\mathbf{27}$ & $\mathbf{184}$ & $\mathbf{241}$ & $\mathbf{306}$ & $\mathbf{611}$ & $\mathbf{0.111}$\\
				\bottomrule
			\end{tabular}
			
			\caption{For all ranked list sizes, \irflname consistently localizes
				more defects (higher $\mathit{hit}$@$k$) and ranks buggy statements
				higher (lower EXAM) than statement-level BLUiR that does not consider
				suspicious file scores when evaluated on the 815~defects available in
				the Defects4J~v2.0.}
			\label{fig:bluesvsbluir}
		\end{center}
	\end{figure}
	
	\subsubsection{\irflname vs.\ Baseline}
	\label{sec:blues_vs_baseline}
	
	We implement a version of statement-level BLUiR~(vanilla BLUiR) that 
	does not consider the suspiciousness scores of the ranked suspicious 
	files and instead ranks the suspicious statements only based on their similarity 
	to the bug reports.
	Figure~\ref{fig:bluesvsbluir} compares \irflname' and vanilla BLUiR performance
	on the 815 defects. For all list sizes, \irflname consistently outperforms vanilla BLUiR
	with higher $\mathit{hit}$@$k$ and lower EXAM.
	
	\begin{tcolorbox}
		For APR-relevant scenarios ($k \geq 25$), \irflname consistently localizes
		more defects and ranks buggy statements higher than the state-of-the-art,
		\emph{supervised}, statement-level IRFL technique used in iFixR. \irflname
		also consistently outperforms a statement-level baseline that ignores
		suspicious files' ranks. \textbf{(RQ1)}
	\end{tcolorbox}
	
	\subsection{\sbir's Evaluation}
	\label{sec:sbir_evalution}
	
	We next compare \sbir with its underlying SBFL and \irflname
	(Section~\ref{sec:sbir_results}) and with state-of-the-art FL techniques
	(Section~\ref{sec:sbir_vs_soa}). As \sbir's ranked lists are at most
	100~statements, our comparisons use that maximum.
	To account for the randomness in \sbir's Monte Carlo algorithm, 
	we compute \sbir using 10 random seeds, reporting the mean, standard
	deviation (stdev), and coefficient of variation~($\text{cv} =
	\frac{\text{stdev}}{\text{mean}}$, which measures variability
	in relation to the mean of the population). 
	A coefficient of variation less than 0.1 means the 10~seeds' results are 
	tightly coupled~\cite{Aronhime14}.
	
	\subsubsection{\sbir's FL Performance~(\textbf{RQ2})}
	\label{sec:sbir_results}
	
	Figure~\ref{fig:sbir_results} shows the FL performance of \sbir, SBFL, and
	\irflname for different list sizes. \sbir consistently localizes more defects
	(higher $\mathit{hit}$@$k$) and ranks buggy statements higher (lower EXAM) than SBFL
	and \irflname. For example, considering top-100 statements, \sbir, on average, localizes
	8 more defects than SBFL and 251 more defects than \irflname. Comparing the
	ranks of buggy statements in the top-100 ranked lists, \sbir, on average,
	ranks buggy statements 19 (EXAM 0.187) while SBFL 22 (EXAM 0.220) and
	\irflname 27 (EXAM 0.270). These results confirm prior findings suggesting
	that combining FL techniques can lead to better FL~\cite{Li19, Zou19,
		Jiang19ase, Sohn17, Le16issta, Xuan14icsme}. Thus, an APR tool using \sbir
	gets earlier opportunities to patch the buggy statements and a more diverse
	set of localized defects than using SBFL or \irflname.
	
	\begin{figure}[t]
		\begin{center}
			\small
			\resizebox{\columnwidth}{!}{
				\begin{tabular}{lr|cccc|ccc}
					\toprule
					\multicolumn{2}{l|}{(815 defects)} & \multicolumn{4}{c|}{$\mathit{hit}$@$k$} & \multicolumn{3}{c}{EXAM} \\
					\midrule
					& & $k=$ 1  & 25 & \phantom{0}50  & 100 &  $k=$ 25 & \phantom{0}50  & 100 \\
					\midrule
					\multicolumn{2}{l|}{SBFL}      & \phantom{0}88 & 408 & 475 & 549 & 0.287 & 0.240 & 0.220 \\
					\multicolumn{2}{l|}{\irflname} & \phantom{0}27 & 184 & 241 & 306 & 0.332 & 0.300 & 0.270 \\
					\midrule
					\midrule
					\textbf{\sbir} & mean      & $\mathbf{101}$    &	$\mathbf{419}$   &	$\mathbf{489}$ & $\mathbf{557}$ &	$\mathbf{0.256}$  & $\mathbf{0.215}$	& $\mathbf{0.187}$ \\
					(10 seeds)	& stdev    & $7.60$ &	$5.01$	& $5.40$ & $4.22$ & $0.006$ &	$0.006$ &	$0.005$ \\
					& cv		& $0.08$ &	$0.01$ &	$0.01$	& $0.01$	& $0.023$ &	$0.026$ & $0.028$ \\
					\bottomrule
				\end{tabular}
			}
			\caption{Comparing \sbir, SBFL, and \irflname FL performance on the 815~defects in Defects4J~(v2.0). 
				For all list sizes, \sbir consistently localizes more
				defects (higher $\mathit{hit}$@$k$) and places buggy statements higher in the list (lower EXAM) 
				than underlying SBFL and \irflname.  
			}
			\label{fig:sbir_results}
		\end{center}
	\end{figure}
	
	\begin{tcolorbox}
		For all list sizes we consider, \sbir consistently localizes more defects and
		ranks buggy statements higher than underlying SBFL and \irflname.
		\textbf{(RQ2)}	
	\end{tcolorbox}
	
	\subsubsection{\sbir vs.\ State of the Art~(\textbf{RQ3})}
	\label{sec:sbir_vs_soa}
	
	We compare \sbir to 9~standalone FL techniques and a supervised learning-to-rank
	approach~\cite{Kuo14} used by existing combining FL techniques.
	
	\looseness-1
	\textbf{\sbir vs.\ Standalone FL.}
	Our evaluation considers techniques that were previously evaluated on Defects4J, make no
	assumptions about a priori knowing the buggy file, and localize buggy
	statements (as opposed to methods or files).
	We compare \sbir with  9 such standalone FL techniques used in a recent FL
	evaluation~\cite{Zou19}: 
	two SBFL\,---\,Ochiai and DStar; 
	two mutation-based FL (MBFL)\,---\,Metallaxis and MUSE; 
	three slicing\,---\,union, intersection, and frequency;
	one stack trace FL; 
	and one predicate switching FL.
	The existing evaluation~\cite{combineFL} provides a dataset of the
	357~defects of Defects4J~(v1.0) annotated with suspiciousness scores of the
	9~techniques, but does not release the implementations of the individual
	techniques. We recreate ranked lists of the 9~techniques from the dataset.
	334 of the 357~defects have bug reports available, making \sbir possible. We
	use these 334 defects for our analysis.
	Figure~\ref{fig:sbirvsstandalone} compares the
	9~techniques with \sbir. For all
	list sizes, \sbir consistently localizes more defects (higher $\mathit{hit}$@$k$) and
	ranks buggy statements higher (lower EXAM) than all of the 9 prior techniques. 
	
	\begin{figure}[t]
		\begin{center}
			\resizebox{\columnwidth}{!}{%
				\begin{tabular}{lr|cccc|c}
					\toprule
					(334 defects) & & \multicolumn{4}{c}{$\mathit{hit}$@$k$} & EXAM \\
					\midrule
					family & technique &  $k=$ 1 & \phantom{0}25 & \phantom{0}50 & 100 & $k = 100$ \\
					\midrule
					\multirow{3}{*}{SBFL} & Ochiai & 30 & 168 & 196 & 221 & 0.254 \\	
					& DStar  & 32 & 169 & 199 & 222 & 0.254 \\ 
					\midrule
					\multirow{2}{*}{MBFL} & Metallaxis & 40 & 154 & 175 & 195 & 0.238 \\	
					& MUSE  & 26 & \phantom{0}96 & 104 & 118 & 0.193 \\ 
					\midrule
					\multirow{3}{*}{slicing} & slicing-union & 21 & \phantom{0}87 & 100 & 111 & 0.462 \\	
					& slicing-intersection  & 18 & \phantom{0}71 & \phantom{0}81 & \phantom{0}91 & 0.481 \\ 
					& slicing-frequency  & 21 & \phantom{0}86 & 100 & 112 & 0.458 \\
					\midrule
					stack trace & stack trace & 16 & \phantom{0}28 & \phantom{0}28 & \phantom{0}28 & 0.663 \\	
					\midrule
					predicate switching & predicate switching & \phantom{0}9 & \phantom{0}24 & \phantom{0}24 & \phantom{0}24 & 0.662 \\
					\midrule
					\midrule
					\textbf{\sbir} (10 seeds) & mean   & $\mathbf{48}$ & $\mathbf{177}$ & $\mathbf{207}$ & $\mathbf{231}$ & $\mathbf{0.175}$ \\
					& stdev  & $4.31$ &  $4.16$	& $2.92$ &	$2.32$ &	$0.006$ \\
					& cv &  $0.09$ &	$0.02$ & 	$0.01$	& $0.01$ &	$0.034$ \\
					\bottomrule 
				\end{tabular}
			}
			\caption{Comparing \sbir to 9 standalone
				FL~techniques on 334 defects from Defects4J~(v1.0). For all list sizes,
				\sbir consistently localizes more defects (higher $\mathit{hit}$@$k$) and
				places buggy statements higher in the ranked lists (lower EXAM) than each of
				the 9 techniques. 
			}
			\label{fig:sbirvsstandalone}
		\end{center}
	\end{figure}
	
	\textbf{\sbir vs.\ Supervised Combining FL Techniques.} 
	Supervised learning-to-rank approaches (e.g., RankSVM~\cite{Kuo14},
	RankBoost~\cite{Freund03}, RankNet~\cite{Burges05}, FRank~\cite{Tsai07},
	LambdaRank~\cite{Burges06}) can combine FL techniques.
	Most such state-of-the-art techniques (e.g., CombineFL~\cite{Zou19}, 
	Fluccs~\cite{Sohn17}, TraPT~\cite{Li17}, Savant~\cite{Le16issta}) use
	RankSVM~\cite{Kuo14}.
	Thus, we compare our unsupervised \sysname with supervised RankSVM in
	combining SBFL and \irflname.\footnote{We could not compare \sysname
		to the deep learning-based DeepFL~\cite{Li19} because DeepFL's data is not public (\url{https://github.com/DeepFL/DeepFaultLocalization/issues/4}).} 
	We first create a dataset of the 815~defects by
	annotating program statements of each defect with normalized suspiciousness
	scores obtained using our SBFL and \irflname, along with the ground truth
	information. We then use this annotated dataset to train the RankSVM model using
	SBFL's and \irflname' scores as features. To evaluate
	the trained model, we use the CombineFL framework~\cite{combineFL} that uses
	10--fold cross validation and computes $E_{\mathit{inspect}}$@$k$ and EXAM
	metrics. The $E_{\mathit{inspect}}$@$k$ metric break ties by computing the
	expected rank of buggy statement in the ranked lists and then counts the
	number of defects whose buggy statements have expected rank $\leq k$. 
	(As there are no ties in \sbir lists, $E_{\mathit{inspect}}$@$k$ is the same 
	as the $\mathit{hit}$@$k$ for \sbir.)
	The EXAM scores are computed using the expected ranks of buggy statements in the
	lists therefore, we denote it as EXAM$_{\mathit{inspect}}$.
	Figure~\ref{fig:raflvssvm} compares \sbir (implemented using \sysname as
	described in Section~\ref{sec:sbir}) and SBIR~(RankSVM) (the combination of
	SBFL and \irflname, combined using RankSVM).
	For all lists of sizes, \sbir consistently localizes significantly 
	more defects (higher $E_{\mathit{inspect}}$@$k$) and
	ranks buggy statements higher (lower EXAM$_{\mathit{inspect}}$) than RankSVM.
	The fact that \sbir is unsupervised and requires no training data is a
	further advantage over the supervised RankSVM approach.
	
	\begin{figure}[t]
		\centering
		\resizebox{\columnwidth}{!}{
			\begin{tabular}{lr|cccc|c}
				\toprule
				\multicolumn{2}{l|}{(815 defects)} & \multicolumn{4}{c}{$E_{\mathit{inspect}}$@$k$} & EXAM$_{\mathit{inspect}}$ \\
				\midrule
				\multicolumn{2}{l|}{technique} &  $k=$ 1 & 25 & \phantom{0}50  & 100 & $k = 100$\\
				\midrule
				\multicolumn{2}{l|}{SBIR (RankSVM)} & \phantom{0}50 & 270 & 328 & 396 & $0.236$ \\	
				\midrule
				\midrule
				\textbf{\sbir(RAFL)} & mean      & $\mathbf{101}$   &	$\mathbf{419}$   &	$\mathbf{489}$ & $\mathbf{556}$ &	$\mathbf{0.187}$ \\
				(10 seeds) & stdev	  & $7.60$	& 	$5.01$	& 	$5.41$ & $4.22$ &	$0.005$ \\
				& cv		  & $0.08$	& 	$0.01$	& 	$0.01$ & $0.01$ & $0.027$ \\
				\bottomrule
			\end{tabular}
		}
		\caption {Comparing \sbir to a supervised-RankSVM combination of SBFL and
			\irflname on 815 defects from Defects4J~(v2.0). For all list sizes, \sbir
			consistently localizes more defects (higher $E_{\mathit{inspect}}$@$k$) and
			places buggy statements higher in the ranked lists (lower EXAM) than
			the RankSVM combination. 
		}
		\label{fig:raflvssvm}
	\end{figure}

\begin{tcolorbox}
	\looseness-1
	\sbir outperforms 9 standalone FL techniques and a supervised technique used
	by existing combiners.
	\textbf{(RQ3)} 
\end{tcolorbox}

\subsection{APR Evaluation Tools, Dataset, and Metrics}
\label{sec:APR Evaluation Tools, Dataset, and Metrics}

We ran our experiments evaluating FL's effect on APR using a cluster of
50~compute nodes, each with a Xeon E5-2680 v4 CPU with 28~cores
(2~processors, 14~cores each) running at 2.40GHz. Each node had 128GB of RAM
and 200GB of local SSD. We launched multiple repair attempts in parallel,
each requesting 4 cores on one compute node. 
We next describe the APR tools we evaluate, our dataset, and the metrics used.

\subsubsection{APR Tools Evaluated}
\label{sec:apr_using_sbir}

Instead of developing a new APR tool or arbitrarily selecting tools
from state-of-the-art, we select Arja~\cite{Yuan20} and SimFix~\cite{Jiang18}
that are the most~($\mathit{Sen} = 66.9\%$) and least~($\mathit{Sen} = 29.5\%$) FL-sensitive 
general purpose repair tools out of the 11~APR tools evaluated in a recent 
study~\cite{Liu21} for their FL sensitivity. 
We select a third tool, SequenceR~\cite{Chen19}, which uses fundamentally different
repair approach than Arja and SimFix, and whose FL-sensitivity~($\mathit{Sen} = 39.5\%$) 
lies between Arja and SimFix.   
Our tool selection criteria require that tools apply to general defects, 
rather than specialized, and have public implementations available so that they can 
be customized to take precomputed FL results.
Arja, SequenceR, and SimFix use genetic-programming-\cite{Koza92}, 
neural-machine-translation-\cite{Tufano19}, and fix-pattern-mining-based~\cite{Liu19issta} 
repair approaches, respectively.
Although there are more effective learning-based APR tools (e.g., CURE~\cite{Jiang21}) than
SequenceR, which is only applicable to single-line-edit defects, we use SequenceR
because its implementation is public and can be customized.

Using the dataset described next (Section~\ref{sec:apr_dataset}), we use Arja and
SimFix to repair 689~single-file-edit defects and SequenceR to repair
129~single-line-edit defects using SBFL, \irflname, and \sbir for FL. We use
the developer-written tests to validate the produced patches.
As \sbir's ranked lists contain at most 100 suspicious statements, 
to fairly evaluate repair tools with respect to all three FL techniques, 
we limit repair tools to use top-100 suspicious program statements obtained by the three FL techniques.
The original SequenceR evaluation~\cite{Chen19} used (manually created) perfect FL and top-10 statements to repair a defect.
We do the same for SequenceR.
We do not otherwise modify the implementations of the three repair tools 
except customizing them to use our precomputed FL results. 

\subsubsection{Dataset}
\label{sec:apr_dataset}

Manually assessing the correctness of patches
that modify multiple files is error-prone and suffers from bias~\cite{Le19, Ye21}. 
To reduce errors and bias, we consider the 689~single-file-edit defects from the 815~defects from
Section~\ref{sec:FL Evaluation Dataset and Metrics}.
As SequenceR applies to single-line-edit defects, 
we use the 129~single-line-edit defects that are 
a subset of the 689~defects.
Figure~\ref{fig:d4j} shows the distribution
of the 689~single-file-edit defects and 129~single-line-edit defects 
across the 17~projects in the Defects4J benchmark.

\subsubsection{Metrics}
\label{sec:apr_metrics}

Prior repair tools' evaluations that measure patch correctness use either
manual inspection~\cite{Martinez17, Xiong18, Le19} or automatically-generated
evaluation test suites~\cite{Xin17, Xiong17, Le19, Motwani22, Afzal21}. While
manual inspection is subjective and could be biased, using evaluation
test-suites could inaccurately measure patch correctness~\cite{Le19}.
Therefore, we propose a novel patch evaluation methodology that uses a hybrid
of these methods to evaluate patch correctness.

For each patched defect, we use the developer-patched program (available for
all Defects4J defects) as an oracle and use EvoSuite~\cite{Fraser13} to
generate 10~held-out test suites using 10~seeds, a search budget of 12~minutes per
seed, and a coverage criterion of maximizing line coverage of the
developer-modified classes. 
We use EvoSuite because it is typically used to generate tests 
for regression oracles, and because prior studies~\cite{Lima21, Motwani22} preferred EvoSuite 
for this task. Most studies using EvoSuite use a 3~minute budget per seed, 
but using longer time budgets leads to better quality tests~\cite{Motwani22}. 
Therefore, we used 12~minutes (4 times what most prior studies use) per seed, for 10~seeds. 

To check the correctness of
an automatically produced patch, we first execute the held-out evaluation
tests on the patch. If any test fails, we annotate such patch as
\emph{plausible} (the term used for a patch that passes developer-written
tests but is incorrect~\cite{Qi15}). This methodology is the state-of-the-art
objective (but potentially incomplete~\cite{Le19}) automated test-driven
patch correctness methodology~\cite{Motwani22}. If all the evaluation tests pass, we manually compare the patch
against the developer's patch. If the patch is semantically equivalent to the
developer's patch, we annotate it as \emph{correct}. If it is not, we
annotate it as \emph{plausible}. If a patch is partially correct or we cannot
determine its semantic equivalence because it requires extensive domain
knowledge, which happens when the modifications are made to methods that are
different from developer-modified ones, we conservatively annotate it as
\emph{plausible}. Thus, our patch evaluation methodology is conservative as
it only considers a patch to be \emph{correct} if it passes all held-out
evaluation tests and is also semantically equivalent to the developer's
patch. To study the effect of improving FL on APR, we compare the number of
defects a repair tool correctly patches~(\emph{repair quality}), using different FL techniques.
Since we had run \sbir with 10 seeds, we executed the APR tool experiments
ten times, once for each \sbir result.
We verified that the defects patched in each run is representative of all 10,
but we only manually analyzed the patches for correctness for one run because
of the significant manual effort involved.

\subsection{Effect of \sbir on APR quality~(\textbf{RQ4})}
\label{sec:apr_results}

\begin{figure}[t]
	\resizebox{\columnwidth}{!}{%
		\begin{tabular}{rccc}
			& \includegraphics[scale=0.33]{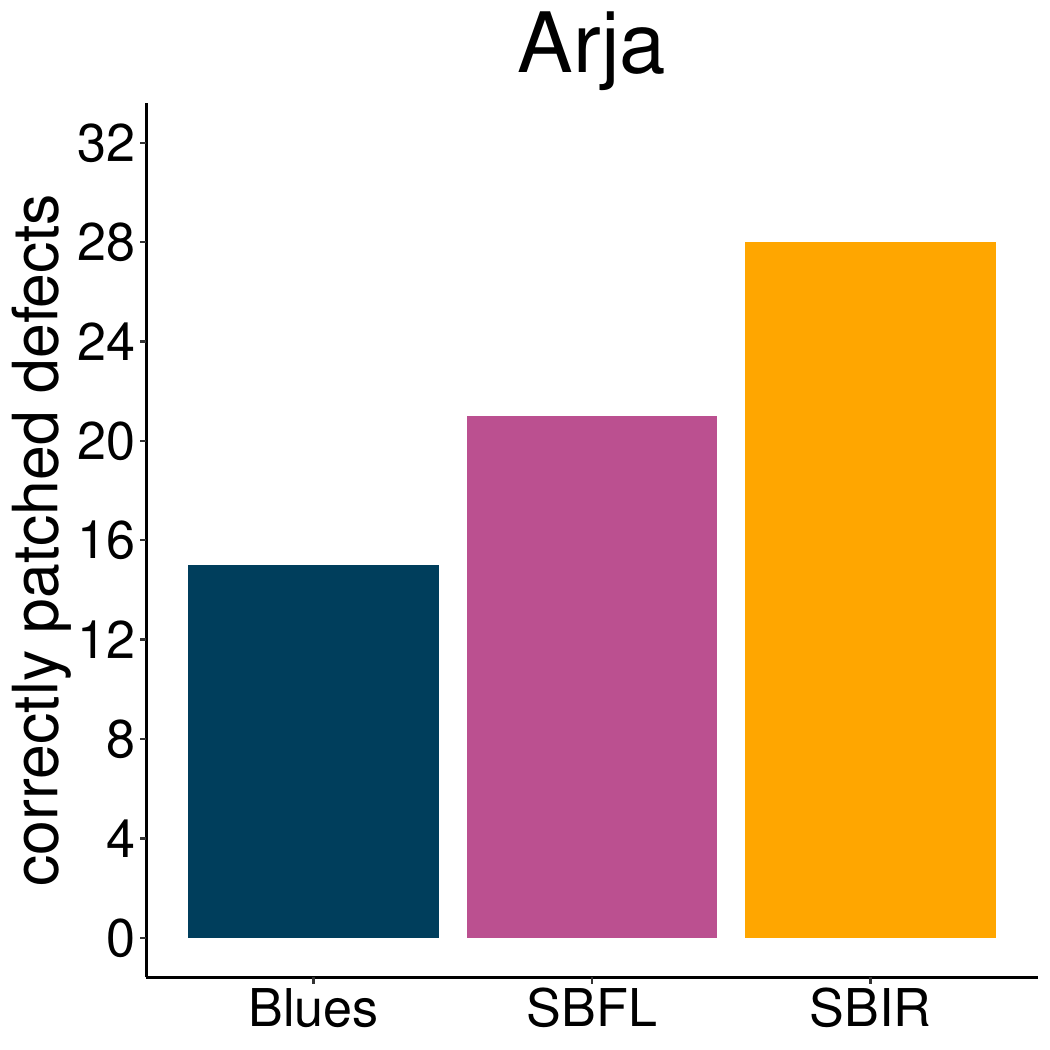} & \includegraphics[scale=0.33]{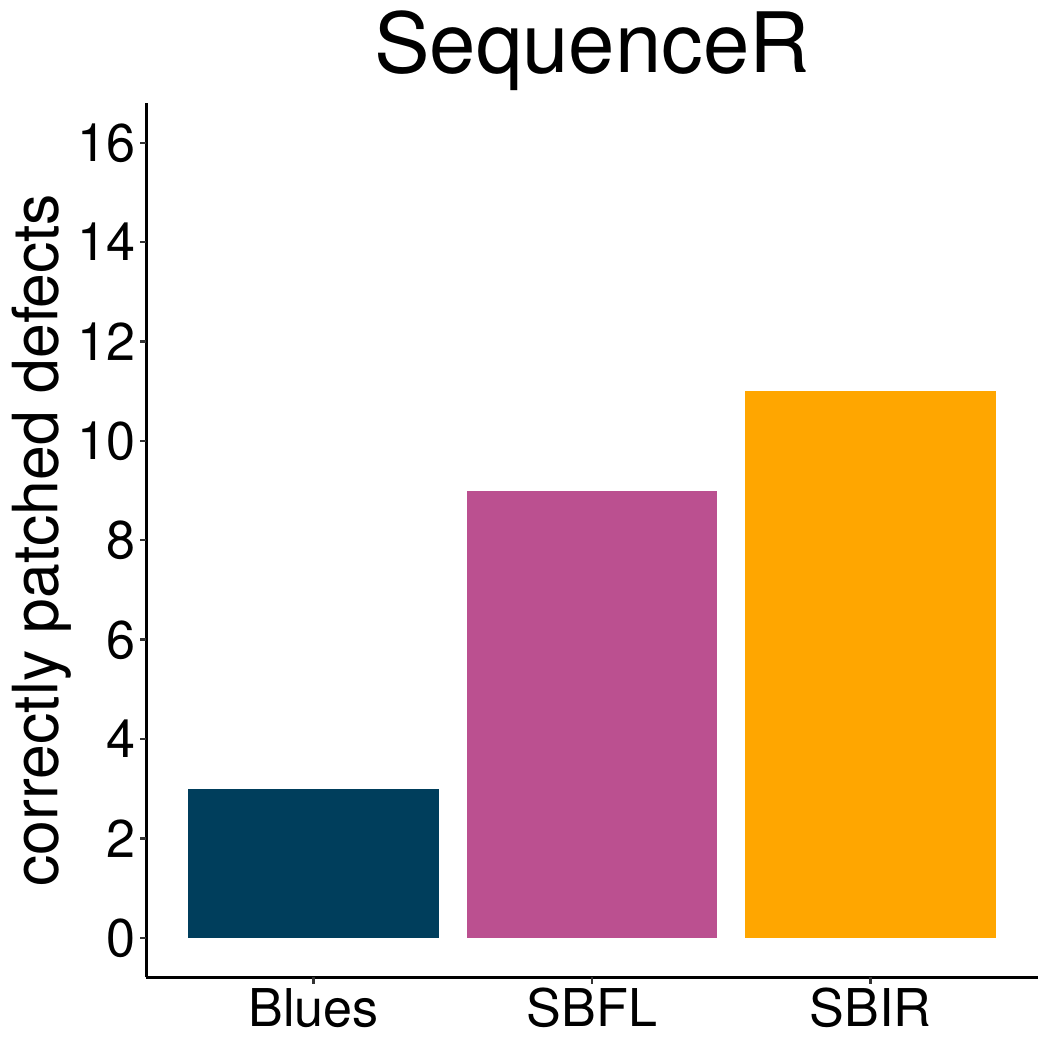} & \includegraphics[scale=0.33]{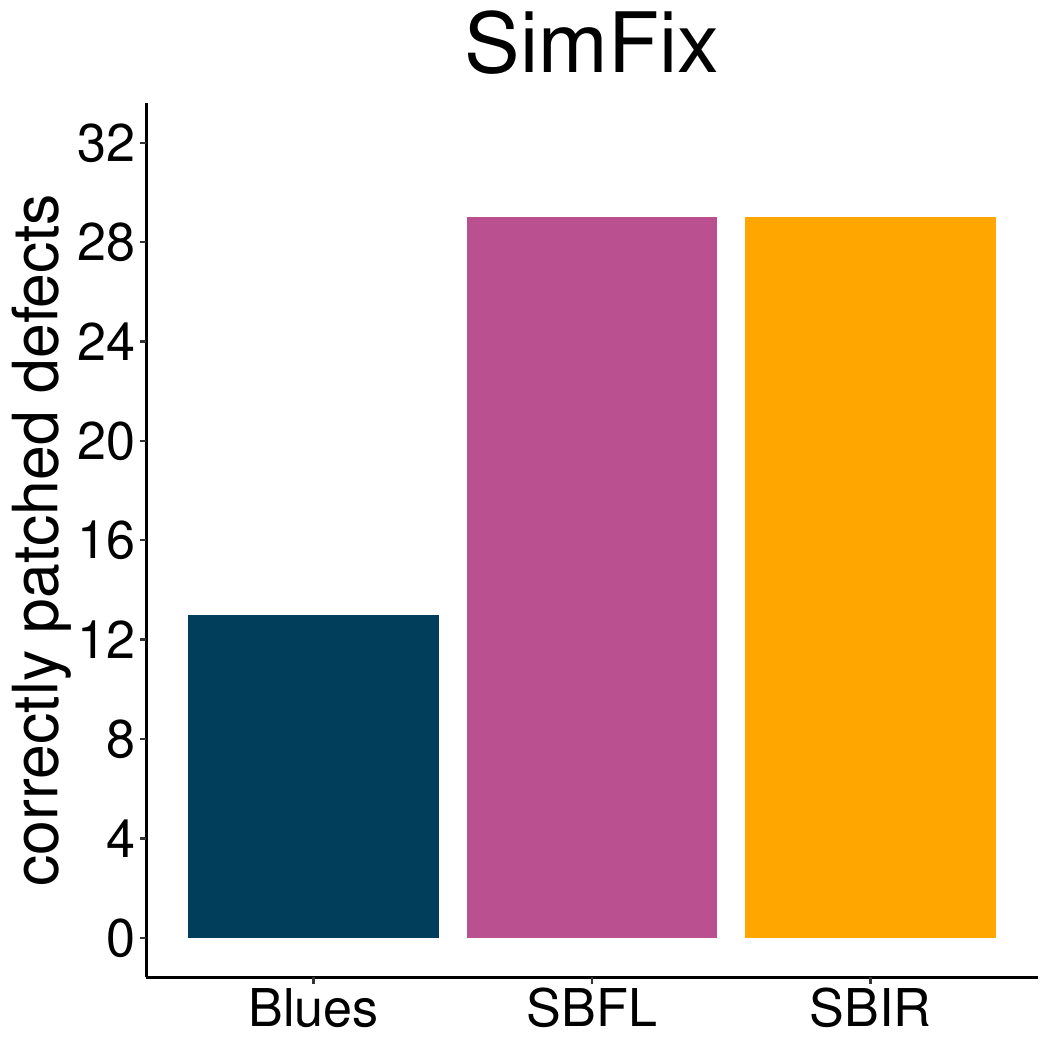} \\
		\end{tabular}
	}	
	\resizebox{\columnwidth}{!}{%
		\begin{tabular}{lccc}	
			\toprule
			& 		Arja & SequenceR & SimFix \\
			& (689~defects) & (129~defects) & (689~defects) \\
			\bottomrule
			\rowcolor{lightgray} \multicolumn{4}{c}{repair quality assessment} \\
			SBFL & 21 & 10 & 30 \\
			\irflname & 15 & \phantom{0}4 & 13\\
			SBFL $\cup$ \irflname & 25 & 12 & 30 \\
			\midrule
			\textbf{\sbir} & $\mathbf{28}$ & $\mathbf{12}$ & $\mathbf{30}$ \\
			\bottomrule		
			\rowcolor{lightgray}
			\multicolumn{4}{c}{localization error assessment} \\
			perfect FL & 21 & 24 & 21 \\
			upper bound & 36 & 24 & 32 \\
			\midrule
			\multicolumn{4}{c}{$\downarrow$ \# of defects not correctly patched due to localization error $\downarrow$} \\
			\midrule
			SBFL    & $15$ & $14$  & $\phantom{0}2$ \\
			\irflname       & $21$ & $20$  & $19$  \\
			\midrule
			\textbf{\sbir}  & $\mathbf{\phantom{0}8}$  & $\mathbf{12}$  &
			$\mathbf{\phantom{0}2}$ \\
			\bottomrule
		\end{tabular}
	}
	\caption{\sbir improves repair quality and reduces localization errors
		for more FL-sensitive APR tools.   	
		Arja and SequenceR, more FL-sensitive tools, correctly patch complementary 
		defects using SBFL and \irflname, and benefit more from using \sbir. 
		SimFix, a less FL-sensitive repair tool, correctly patches the same number 
		of defects using \sbir as SBFL but more than \irflname.}
	\label{fig:repairresults_overall}
\end{figure}

The top of Figure~\ref{fig:repairresults_overall} compares repair quality of
the three repair tools using the three FL techniques. Arja and
SequenceR correctly patch more defects when using \sbir than when using SBFL
or \irflname. Specifically, Arja using SBIR correctly repairs 7~(33\%) more
defects than using SBFL and 13~(87\%) more defects than using \irflname.
SequenceR using SBIR correctly patches 2~(20\%) (out of a smaller subset of
single-line defects) more defects than using SBFL and 8~(200\%) more
defects than using \irflname. SimFix unsurprisingly correctly patches the
same number of defects when using SBFL but 17~(131\%) more defects than
using \irflname. More FL-sensitive repair tools, Arja and SequenceR,
correctly patch complementary defects using SBFL and \irflname, as evident by
the row showing the union of defects they patch using SBFL and \irflname.
However, as the less FL-sensitive SimFix uses test case
purification~\cite{Xuan14} and expands each suspicious statement by
$\pm 5$ lines to address inaccurate FL, it does not patch complementary
defects.

\textbf{Localization Error Analysis.}
Multiple factors can prevent repair tools from producing correct patches. For
example, 
if inaccurate FL ranks irrelevant non-buggy statements as more suspicious
than buggy statements, the tool may produce plausible patches before having a
chance to explore the buggy statement. This phenomenon is called APR
localization error~\cite{Koyuncu19}. We next measure SBFL's, \irflname',
and \sbir's effect on localization error. We execute each of the three repair
tools using perfect (manual) FL and measure the number of correctly patched
defects (``perfect FL'' row in Figure~\ref{fig:repairresults_overall}). We
compute the ``upper bound''~\cite{Liu21} number of defects a repair tool can
correctly patch as the union of defects correctly patched using the perfect
FL and our three FL techniques. (Note that Arja and SimFix consider multiple
suspicious statements and can patch more defects using \sbir than perfect FL.
Their repair algorithms fail to construct patches for some defects when FL's
ranked lists do not contain certain non-buggy statements adjacent to the
buggy ones.) We then compute the localization error for
each FL technique: the difference between the upper bound and the number of
defects correctly patched using that FL technique. The bottom three rows of
Figure~\ref{fig:repairresults_overall} show that using \sbir significantly
reduces the number of defects not patched due to localization error for the
more FL-sensitive repair tools, compared to SBFL and \irflname.

\begin{figure}[t]
	\footnotesize
	\begin{lstlisting}[linewidth=\columnwidth, firstnumber=709, basicstyle=\tiny, xleftmargin=1cm]
		case Token.MOD:
		if (rval == 0) { 
			error(DiagnosticType.error("JSC_DIVIDE_BY_0_ERROR", "Divide by 0"), right);    // Blues(38) SBIR(40)
			return null;
		}    
		result = lval % rval;
		break;
		case Token.DIV:
		if (rval == 0) { 
			error(DiagnosticType.error("JSC_DIVIDE_BY_0_ERROR", "Divide by 0"), right);    // SBFL(1) Blues(36) SBIR(1)
			return null;
		}    
	\end{lstlisting}
	\caption{The two non-consecutive buggy statements (lines 711 and 718) that 
		cause the Closure-78 defect. The annotations show which of the three FL techniques 
		localize the buggy statements and their ranks in the respective lists.}
	\label{fig:closure78}
\end{figure}

Overall, Arja and SequenceR significantly benefit from \sbir. Arja using
\sbir correctly patches 28~(78\%) of the 36~upper bound defects, whereas
using SBFL, it only patches 21~(58\%) and using \irflname only 15~(42\%).
SequenceR using \sbir correctly patches 12~(50\%) of the 24~upper bound
defects, whereas using SBFL, it patches 10~(42\%) and using \irflname
4~(17\%). SimFix, correctly patches 30~(94\%) of the 32~upper bound defects
using both \sbir and SBFL, and patches 13~(41\%) using \irflname.

\looseness-1
\textbf{Case Study Illustrating How \sbir Helps APR.} 
The three APR tools using \sbir correctly patched 7~defects
(Chart-12, Closure-68, Closure-78, Closure-86, Closure-92, Lang-10, and
Lang-20) in Defects4J that none of the existing 14~APR tools
patch.
That is a 7.5\% improvement over the 93 
defects in Defects4J~(v1.0) that at least one of the 14 tools correctly
patches~\cite{Liu19}.
Closure-78 and Lang-20 require editing multiple code locations in a single
file, and most repair tools struggle to patch these kind of defects.
For example, Arja using \sbir correctly patches
Closure-78, whose repair involves deleting two non-consecutive statements
(lines\,711~and~718 in Figure~\ref{fig:closure78}) to fix a division-by-0
error. None of the existing 14~APR tools~\cite{Liu19}, nor Arja with our SBFL
or \irflname, patch this defect. For this defect, SBFL ranks only line\,718
in the 1st position; \irflname ranks line\,718 36th and line\,711 38th; and
\sbir ranks line\,718 1st and line\,711 40th. Arja using SBFL produces a
plausible, but only partially correct patch that deletes line\,718 while Arja
using \irflname does not produce a patch, timing out trying to modify the
35~non-buggy statements ranked higher. Arja using \sbir produces a correct
patch (identical to the developer patch) because it finds the buggy statement
at the top of \sbir's list, and then fetches the second buggy statement
because it is also in \sbir's list and because it uses the same variables and
methods as the top-ranked line (Arja's \emph{ingredient screening}
step~\cite{Yuan20}). Arja constructs a correct patch by deleting both the
buggy statements. Thus, it is precisely the \emph{combination} of the
information from bug reports and test executions that enables Arja to
localize and correctly repair this defect.

\textbf{\sbir vs.\ Union of SBFL and \irflname.} 
Since APR using SBFL and IRFL often repairs complementary
defects~\cite{Koyuncu19}, we set out to measure how defects repaired with
\sbir compare to the union of the defects repaired with \irflname and SBFL.
We find that while there is some complementarity, for Arja, \sbir repairs
more defects than the union, suggesting that combining bug reports and tests
not only captures most (though not all) of the benefits of the two, it is
also able to extract a combined benefit where neither \irflname nor SBFL alone
leads to a repair.
Arja patches 25~defects (row~3 in
Figure~\ref{fig:repairresults_overall}) using SBFL and \irflname, including
4~defects (Compress~27, Jsoup~33, Jsoup~55, and Time~15) that Arja could not
patch using \sbir. However, Arja using \sbir patches 28~defects, including
7~defects (Closure~78, Gson~7, Jsoup~39, Jsoup~68, Jsoup~85, JxPath~5, and
Lang~7) that Arja could not patch using SBFL or \irflname. Thus, for Arja,
\sbir is even more beneficial than using both SBFL and \irflname.
SequenceR patches 12~defects using both SBFL and \irflname that include one
defect (Cli~40) that SequenceR could not patch using \sbir. However,
SequenceR using \sbir also patches 12~defects that include one defect
(JacksonCore~25) that SequenceR could not patch using SBFL or \irflname.
Thus, \sbir provides the same benefit to SequenceR as using both SBFL and
\irflname.
SimFix correctly patches the same 30~defects using \sbir as it does using
both SBFL and \irflname. Thus, for SimFix, using \sbir provides the same
benefit as using just SBFL or both SBFL and \irflname.

\looseness-1
\textbf{\sbir vs.\ Original
	Published APR Versions.} 
We find that the three repair tools using \sbir correctly repair somewhat
complementary defects to those the original published versions repaired.
Arja using \sbir correctly patches 4~defects 
(Lang-7, Lang-10, Lang-59, and Math-35) original Arja did not. 
Of the defects in our dataset, the original Arja correctly patched 15
defects~\cite{Yuan20} (plus 3~others that either had no bug reports
(Chart-3) or were multi-file-edit defects (Math-22 and
Math-98)). Of these 15, Arja using \sbir correctly patches 12, but not the other 
3 (Lang-35, Math-39, and Math-86). 
We examined the original evaluation's 
patches\footnote{\url{https://github.com/yyxhdy/defects4j-patches/tree/master/Arja}}  
and found that for these 3~defects, Arja had produced only a single
patch, which is highly uncommon for Arja (it produced many patches
for all other defects it patched), suggesting that there is something special
about these defects or the process the Arja evaluation followed in repairing
them. Overall, Arja with \sbir correctly patches 1 more defect than the original
Arja.
SimFix using \sbir correctly patches 3~defects (Closure-68, Closure~92, and
Closure-126) original SimFix did not. Of the defects in our dataset, original
SimFix correctly patched 21~defects~\cite{Jiang18} (plus 6 others that either had
no bug reports (Chart-3, Chart-7, Chart-20) or were multi-file-edit defects 
(Closure-63, Math-71, and Math-98)). Of these 21,
SimFix with \sbir correctly patches 19. (Note that the original
evaluation~\cite{Jiang18} listed 7 more defects (Closure-115, Lang-16,
Lang-27, Lang-39, Lang-41, Lang-50, and Lang-60) as patched correctly. The authors 
subsequently identified one of those (Lang-27) as 
incorrect,\footnote{\url{https://github.com/xgdsmileboy/SimFix/tree/master/final/result}}
and our analysis revealed that the six others are also incorrect.
SimFix with \sbir could not patch the remaining two defects (Math-35 and
Math-63). Overall, SimFix with \sbir correctly patches 1 more defect than the
original SimFix.
SequenceR's original evaluation used perfect FL~\cite{Chen19}, so a
direct comparison is not appropriate. With perfect FL, 
original SequenceR patched 14~defects correctly, and with \sbir, 
it patches 6 (Chart-11, Closure-73, Closure-86, Lang-59, Math-58, and Math-75) 
of those 14.

\begin{tcolorbox}	
\sbir significantly improves repair quality and reduces localization errors
for more FL-sensitive APR tools, and enables correctly repairing some defects that 
they cannot repair with other FL techniques. For less FL-sensitive APR, \sbir
provides the same repair quality as SBFL. Using \sbir, we are able to
correctly repair 7 defects never previously automatically repaired correctly
by existing techniques. \textbf{(RQ4)}
\end{tcolorbox}

\subsection{Discussion and Threats to Validity}
\label{sec:discussion}

\looseness-1
Our approach requires a bug report and a bug-exposing test. This requirement
is not always met: several defects in Defects4J~(v2.0) have no documented bug
reports, and prior studies~\cite{Koyuncu19, Just18} show that for 92\% of
defects, bug-exposing tests are added after the bug is reported.
However, most repair tools cannot function without either a failing test or a bug
report, and existing repair tools that use only bug reports are not fully
automatic (a human must validate the proposed patches). 
Meanwhile, while test-driven APR can be fully automated, most patches it
produces are incorrect~\cite{Noda20, Motwani22, Smith15fse}. Our work extends
APR to use \sbir, which uses both bug reports and test suites, enabling
repair tools to be fully automated and to produce higher-quality patches.
This is a worthwhile achievement even if not all defects in industrial
settings have the requisite artifacts, and may motivate developers to create
the artifacts in the future, which reinforces an already recommended practice.
Further, combining the available user inputs to improve 
APR can foster trust in the generated patches~\cite{Noller22}, thereby
helping the adoption of repair techniques.

\irflname' effectiveness depends on the quality 
of bug reports. For example, \irflname could not localize the 
Chart~2 defect\footnote{\url{https://sourceforge.net/p/jfreechart/bugs/959/}} 
because its bug report only contains a URL and no description. 
This caused \sbir to lower the rank of the buggy statement in its ranked list
of suspicious statements.

It is not a goal of our study to develop the best way to combine FL
techniques. Instead, because existing combining techniques are trained on
Defects4J, we could not use them in our evaluation, so we created \sysname,
an unsupervised combining method. We show that \sysname outperforms RankSVM,
a state-of-the-art supervised combining method, and that it is sufficient to
demonstrate improvement in APR performance. However, comparing \sysname with
\emph{all} supervised methods is out of scope of this study.

\looseness-1
Our evaluation aims to measure the impact of combined IRFL and SBFL on APR in
a way that will generalize to a wide range of APR techniques. That is why our
evaluation uses three diverse APR techniques. The design of our study allows
estimating the \sbir's impact on a repair technique based on its
FL-sensitivity. For example, a recent technique Recoder~\cite{Zhu21} has an
FL-sensitivity of 34.5\%, similar to SequenceR's 39.5\%. Thus, we expect
\sbir's impact on Recoder will be similar to that on SequenceR, but smaller
than that on Arja (sensitivity 66.9\%).

Arja and SequenceR are stochastic and results may vary across executions. We
address this threat by using a large-scale dataset. Executing our study is
highly computationally intensive and required eight weeks of wall-clock time
on a 50-node cluster.
To enable others to independently reproduce our results, and reuse our FL
techniques in improving APR, we make all code and data available.

We address threats to internal validity by reusing publicly available
implementations of repair tools instead of reimplementing them. We address
threats to external validity by selecting diverse APR tools and using
Defects4J~(v2.0) that has significantly more projects and defects than
earlier versions.

\section{Related Work}
\label{sec:relatedwork}

\textbf{Improving APR Performance.} 
Program repair tools typically follow a three step process: 
identifying the location of a defect, producing candidate patches, and
validating those patches. The method used for 
each of these steps can significantly affect the tool's success.
To improve APR, researchers have proposed to use different kinds of
FL strategies~\cite{Assiri17, Yang18, Sun18, Koyuncu19, Jiang19, Lou20}, 
patch generation algorithms (e.g., heuristic-based~\cite{LeGoues12b, Long16, Tian17, Wen18, Jiang18, Petke20}, 
constraint-based~\cite{Afzal21, Wang18, Gulwani18, Mechtaev18, Ke15ase}, 
and learning-based~\cite{Chen19, Gupta17, Saha17}), and 
patch validation methodologies~\cite{Wang20, Yang17fse, Yu19, Tian20, Ye21}.
Assuming perfect FL, recent study~\cite{Liu21} shows that modern repair tools can 
patch significantly more defects. However, assuming perfect FL is unrealistic 
and therefore we propose to improve automated FL used in APR.

Recent APR research has used formal constraints derived by program analyzers
instead of test suites~\cite{Ginelli21, Gao21}. These techniques patch
specific families of defects, such as security vulnerabilities and
exception-causing defects, and our approach to improving general-purpose APR
is complementary. 

APR's fundamental challenge is generating fewer incorrect
patches~\cite{Smith15fse, Motwani22, Qi15}. In some domains, e.g., formal
verification, an oracle exists to determine patch
correctness~\cite{First20oopsla, First22icse, Sanchez-Stern22passport,
Agrawal23icse-demo}, overcoming this problem, though better FL can still lead
to the production of more patches.

\textbf{Improving FL.} Techniques to improve FL can be classified 
into two categories. 
The first category is the standalone techniques.
For example, PRoFL~\cite{Lou20} improves SBFL using patch execution results from
APR, PREDFL~\cite{Jiang19ase} uses runtime statistics from statistical debugging to
improve SBFL, PRFL~\cite{Zhang17} uses the PageRank algorithm, XGB-FL~\cite{Yang20} 
uses a classifier to learn the importance of program statements and features, such as 
execution sequence and semantics, UniVal~\cite{kk21} uses execution profiles and 
the success and failure information from program executions, in conjunction with 
statistical inference, and DeepRL4FL~\cite{Li21} formulates FL as pattern recognition and uses code
coverage representation learning to improve SBFL and MBFL techniques.
The second category 
(e.g., CombinedFL~\cite{Zou19}, DeepFL~\cite{Li19}, Fluccs~\cite{Sohn17},
Savant~\cite{Le16issta}, Multric~\cite{Xuan14icsme}, and TraPT~\cite{Li17})
uses learning-to-rank~\cite{Burges05} machine learning approaches 
such as RankSVM~\cite{Kuo14} to combine multiple FL techniques. 
\sysname outperforms RankSVM, the state-of-the-art supervised method.

Property-based testing, e.g., for software fairness~\cite{Galhotra17fse,
Thomas19science, Angell18demo-fse, Brun18fse-nier}, and automated
oracle generation~\cite{Motwani19icse} can synthesize additional tests to
improve FL in ways complementary to our approach. 

\textbf{FL in Program Repair.}
Most repair tools use SBFL implemented using off-the-shelf 
coverage tracking tools and the Ochiai ranking strategy~\cite{LeGoues12b, Long16, 
Tian17, Gupta17, Saha17, Wen18, Jiang18, Wang18, Mechtaev18, Gulwani18, Afzal21, Chen19, Ye22}. 
R2Fix~\cite{Liu13} and iFixR~\cite{Koyuncu19} are the only two IRFL-based repair tools, 
and no prior repair tool uses combined SBFL and IRFL. Although, using
patch-execution results from repair tools to refine FL results can outperform
state-of-the-art SBFL and MBFL techniques~\cite{Lou20}. Recent studies have
shown the effect of using different technologies, assumptions, and
adaptations of test-suite-based FL techniques on the performance of repair
tools~\cite{Afzal21, Liu19, Jiang19, Sun18, Yang18, Wen17, Assiri17}. Often,
APR researchers omit FL tuning used by their repair tools while
presenting repair performance, which leads to bias in comparing performance
of different repair tools~\cite{Liu21, Liu19}. Further, the tuned FL implementations 
are often tightly coupled to the repair tool implementations, which makes it hard
to reuse them for other repair tools. Our FL techniques can
be used to mitigate this bias as they can serve as a plugin by future repair
tools to decouple their FL implementations from their repair algorithm
implementation, as is done in some frameworks, including
JaRFly~\cite{Motwani22}.

\section{Contributions}
\label{sec:contributions}

We have developed \sbir, an FL technique that uses both bug reports and tests
to localize defects, and showed that it helps improve APR quality for FL-sensitive
tools, repairing more defects correctly than by using other FL techniques. 
Along the way, we also created \irflname, the first statement-level,
information-retrieval-based FL that outperforms the state of the art
without needing ground truth data for training, and
\sysname, a novel unsupervised method for combining arbitrary FL
techniques. Our results demonstrate that combining bug reports and tests 
leads to better FL, and enables higher-quality APR. Our findings support further 
research into improving APR by combining bug-report-based and test-based information.

\section*{Data Availability}
\label{sec:data_availability}
All of our data, source code, and documentation to reproduce 
our results are publicly available~\cite{MotwaniSBIRArtifact}.

\section*{Acknowledgments}
This work is supported by the National Science Foundation under
grants no.\ CCF-1763423 and CCF-2210243.

\balance
\bibliographystyle{abbrv}
\bibliography{relatedwork}

\begin{thebibliography}{100}

\bibitem{Abreu07}
R.~Abreu, P.~Zoeteweij, and A.~J.~V. Gemund.
\newblock On the accuracy of spectrum-based fault localization.
\newblock In {\em TAIC-PART}, pages 89--98, 2007.

\bibitem{Afzal21}
A.~Afzal, M.~Motwani, K.~T. Stolee, Y.~Brun, and C.~{Le Goues}.
\newblock {SOSRepair}: {Expressive} semantic search for real-world program
  repair.
\newblock {\em TSE}, 47(10):2162--2181, October 2021.

\bibitem{Agrawal23icse-demo}
A.~Agrawal, E.~First, Z.~Kaufman, T.~Reichel, S.~Zhang, T.~Zhou,
  A.~Sanchez-Stern, T.~Ringer, and Y.~Brun.
\newblock Proofster: {Automated} formal verification.
\newblock In {\em ICSE Demo}, May 2023.

\bibitem{Angell18demo-fse}
R.~Angell, B.~Johnson, Y.~Brun, and A.~Meliou.
\newblock Themis: {Automatically} testing software for discrimination.
\newblock In {\em ESEC/FSE Demo}, pages 871--875, November 2018.

\bibitem{Aronhime14}
S.~Aronhime, C.~Calcagno, G.~H. Jajamovich, H.~A. Dyvorne, P.~Robson,
  D.~Dieterich, M.~I. Fiel, V.~Martel-Laferriere, M.~Chatterji, H.~Rusinek, and
  B.~Taouli.
\newblock {DCE-MRI} of the liver: Effect of linear and nonlinear conversions on
  hepatic perfusion quantification and reproducibility.
\newblock {\em Journal of Magnetic Resonance Imaging}, 40(1):90--98, 2014.

\bibitem{Assiri17}
F.~Y. Assiri and J.~M. Bieman.
\newblock Fault localization for automated program repair: Effectiveness,
  performance, repair correctness.
\newblock {\em Software Quality Journal}, 25(1):171--199, 2017.

\bibitem{Bader19}
J.~Bader, A.~Scott, M.~Pradel, and S.~Chandra.
\newblock Getafix: {Learning} to fix bugs automatically.
\newblock {\em PACMPL OOPSLA}, 3, October 2019.

\bibitem{Brandenburg13Kendall}
F.~J. Brandenburg, A.~Glei{\ss}ner, and A.~Hofmeier.
\newblock Comparing and aggregating partial orders with kendall tau distances.
\newblock {\em Discrete Mathematics, Algorithms and Applications},
  5(02):1360003, 2013.

\bibitem{Brandenburg13}
F.~J. Brandenburg, A.~Glei{\ss}ner, and A.~Hofmeier.
\newblock The nearest neighbor spearman footrule distance for bucket, interval,
  and partial orders.
\newblock {\em Journal of Combinatorial Optimization}, 26(2):310--332, 2013.

\bibitem{Brun18fse-nier}
Y.~Brun and A.~Meliou.
\newblock Software fairness.
\newblock In {\em ESEC/FSE NIER}, 2018.

\bibitem{Burges06}
C.~Burges, R.~Ragno, and Q.~Le.
\newblock Learning to rank with nonsmooth cost functions.
\newblock In {\em NeurIPS}, pages 193--200, 2006.

\bibitem{Burges05}
C.~Burges, T.~Shaked, E.~Renshaw, A.~Lazier, M.~Deeds, N.~Hamilton, and
  G.~Hullender.
\newblock Learning to rank using gradient descent.
\newblock In {\em ICML}, pages 89--96, 2005.

\bibitem{Campos12}
J.~Campos, A.~Riboira, A.~Perez, and R.~Abreu.
\newblock {Gzoltar}: {An} {Eclipse} plug-in for testing and debugging.
\newblock In {\em ASE}, pages 378--381, 2012.

\bibitem{Chen19}
Z.~Chen, S.~J. Kommrusch, M.~Tufano, L.-N. Pouchet, D.~Poshyvanyk, and
  M.~Monperrus.
\newblock Sequencer: {Sequence}-to-sequence learning for end-to-end program
  repair.
\newblock {\em TSE}, 47(9):1943--1959, 2019.

\bibitem{Christou15}
S.~Christou.
\newblock Cobertura code coverage tool.
\newblock \url{https://cobertura.github.io/cobertura/}, 2015.

\bibitem{Debroy13}
V.~Debroy and W.~E. Wong.
\newblock A consensus-based strategy to improve the quality of fault
  localization.
\newblock {\em Software: Practice and Experience}, 43(8):989--1011, 2013.

\bibitem{Deng14}
K.~Deng, S.~Han, K.~J. Li, and J.~S. Liu.
\newblock Bayesian aggregation of order-based rank data.
\newblock {\em JASA}, 109(507):1023--1039, 2014.

\bibitem{Dwork01www}
C.~Dwork, R.~Kumar, M.~Naor, and D.~Sivakumar.
\newblock Rank aggregation methods for the web.
\newblock In {\em WWW}, pages 613--622, 2001.

\bibitem{ASTStmt}
Eclipse {JDT} {API} specification.
\newblock \url{https://ibm.co/3orMarh}.
\newblock [accessed 4-March-2022].

\bibitem{First22icse}
E.~First and Y.~Brun.
\newblock Diversity-driven automated formal verification.
\newblock In {\em ICSE}, pages 749--761, May 2022.

\bibitem{First20oopsla}
E.~First, Y.~Brun, and A.~Guha.
\newblock {TacTok}: {Semantics}-aware proof synthesis.
\newblock {\em PACMPL OOPSLA}, 4:231:1--231:31, November 2020.

\bibitem{Fraser13}
G.~Fraser and A.~Arcuri.
\newblock Whole test suite generation.
\newblock {\em TSE}, 39(2):276--291, February 2013.

\bibitem{Freund03}
Y.~Freund, R.~Iyer, R.~E. Schapire, and Y.~Singer.
\newblock An efficient boosting algorithm for combining preferences.
\newblock {\em JMLR}, 4:933--969, Nov. 2003.

\bibitem{Galhotra17fse}
S.~Galhotra, Y.~Brun, and A.~Meliou.
\newblock Fairness testing: {Testing} software for discrimination.
\newblock In {\em ESEC/FSE}, pages 498--510, September 2017.

\bibitem{Gao21}
X.~Gao, B.~Wang, G.~J. Duck, R.~Ji, Y.~Xiong, and A.~Roychoudhury.
\newblock Beyond tests: {Program} vulnerability repair via crash constraint
  extraction.
\newblock {\em TOSEM}, 30(2):14:1--14:27, Feb. 2021.

\bibitem{Gay20}
G.~Gay and R.~Just.
\newblock {Defects4J} as a challenge case for the search-based software
  engineering community.
\newblock In {\em SSBSE}, pages 255--261, 2020.

\bibitem{Gazzola19}
L.~Gazzola, D.~Micucci, and L.~Mariani.
\newblock Automatic software repair: {A} survey.
\newblock {\em TSE}, 45(01):34--67, 2019.

\bibitem{Ginelli21}
D.~Ginelli, O.~Riganelli, D.~Micucci, and L.~Mariani.
\newblock Exception-driven fault localization for automated program repair.
\newblock In {\em QRS}, 2021.

\bibitem{Gulwani18}
S.~Gulwani, I.~Radi{\v{c}}ek, and F.~Zuleger.
\newblock Automated clustering and program repair for \looseness-1 introductory
  programming assignments.
\newblock In {\em PLDI}, 2018.

\bibitem{Gupta17}
R.~Gupta, S.~Pal, A.~Kanade, and S.~K. Shevade.
\newblock {DeepFix}: {Fixing} common {C} language errors by deep learning.
\newblock In {\em AAAI}, 2017.

\bibitem{Harman01}
M.~Harman and B.~F. Jones.
\newblock Search-based software engineering.
\newblock {\em Information and Software Technology}, 43(14):833--839, 2001.

\bibitem{Harrold2000}
M.~J. Harrold, G.~Rothermel, K.~Sayre, R.~Wu, and L.~Yi.
\newblock An empirical investigation of the relationship between spectra
  differences and regression faults.
\newblock {\em STVR}, 10(3):171--194, 2000.

\bibitem{Hoffmann09}
M.~R. Hoffmann, B.~Janiczak, E.~Mandrikov, and M.~Friedenhagen.
\newblock {JaCoCo} code coverage tool.
\newblock \url{https://www.jacoco.org/jacoco/}, 2009.

\bibitem{Jiang19ase}
J.~Jiang, R.~Wang, Y.~Xiong, X.~Chen, and L.~Zhang.
\newblock Combining spectrum-based fault localization and statistical
  debugging: {A}n empirical study.
\newblock In {\em ASE}, pages 502--514, 2019.

\bibitem{Jiang19}
J.~Jiang, Y.~Xiong, and X.~Xia.
\newblock A manual inspection of {Defects4J} bugs and its implications for
  automatic program repair.
\newblock {\em Science China Information Sciences}, 62(10):200102, 2019.

\bibitem{Jiang18}
J.~Jiang, Y.~Xiong, H.~Zhang, Q.~Gao, and X.~Chen.
\newblock Shaping program repair space with existing patches and similar code.
\newblock In {\em ISSTA}, 2018.

\bibitem{Jiang21}
N.~Jiang, T.~Lutellier, and L.~Tan.
\newblock {CURE}: {Code}-aware neural machine translation for automatic program
  repair.
\newblock In {\em ICSE}, 2021.

\bibitem{Jones05}
J.~A. Jones and M.~J. Harrold.
\newblock Empirical evaluation of the tarantula automatic fault-localization
  technique.
\newblock In {\em ASE}, pages 273--282, 2005.

\bibitem{Just18}
R.~Just, C.~Parnin, I.~Drosos, and M.~D. Ernst.
\newblock Comparing developer-provided to user-provided tests for fault
  localization and automated program repair.
\newblock In {\em ISSTA}, pages 287--297, July 2018.

\bibitem{Ke15ase}
Y.~Ke, K.~T. Stolee, C.~{Le Goues}, and Y.~Brun.
\newblock Repairing programs with semantic code search.
\newblock In {\em ASE}, pages 295--306, November 2015.

\bibitem{Kirbas21}
S.~Kirbas, E.~Windels, O.~McBello, K.~Kells, M.~Pagano, R.~Szalanski,
  V.~Nowack, E.~R. Winter, S.~Counsell, D.~Bowes, T.~Hall, S.~Haraldsson, and
  J.~Woodward.
\newblock On the introduction of automatic program repair in bloomberg.
\newblock {\em Software}, 38(4):43--51, 2021.

\bibitem{Kolde12}
R.~Kolde, S.~Laur, P.~Adler, and J.~Vilo.
\newblock Robust rank aggregation for gene list integration and meta-analysis.
\newblock {\em Bioinformatics}, 28(4):573--580, 2012.

\bibitem{Koyuncu19arxiv}
A.~Koyuncu, T.~F. Bissyand{\'e}, D.~Kim, K.~Liu, J.~Klein, M.~Monperrus, and
  Y.~L. Traon.
\newblock {D\&C}: {A} divide-and-conquer approach to {IR}-based bug
  localization.
\newblock {\em CoRR}, abs/1902.02703, 2019.

\bibitem{Koyuncu19}
A.~Koyuncu, K.~Liu, T.~F. Bissyand{\'e}, D.~Kim, M.~Monperrus, J.~Klein, and
  Y.~L. Traon.
\newblock {iFixR}: {B}ug report driven program repair.
\newblock In {\em ESEC/FSE}, pages 314--325, 2019.

\bibitem{Koza92}
J.~R. Koza.
\newblock {\em Genetic programming: on the programming of computers by means of
  natural selection}, volume~1.
\newblock MIT press, 1992.

\bibitem{Kuo14}
T.-M. Kuo, C.-P. Lee, and C.-J. Lin.
\newblock Large-scale kernel ranksvm.
\newblock In {\em SIAM Intl. Conf. on data mining}, pages 812--820. SIAM, 2014.

\bibitem{kk21}
Y.~K{\"u}ç{\"u}k, T.~A.~D. Henderson, and A.~Podgurski.
\newblock Improving fault localization by integrating value and predicate based
  causal inference techniques.
\newblock In {\em ICSE}, pages 649--660, 2021.

\bibitem{Le16issta}
T.~B. Le, D.~Lo, C.~{Le Goues}, and L.~Grunske.
\newblock A learning-to-rank based fault localization approach using likely
  invariants.
\newblock In {\em Intl. Symposium on Software Testing and Analysis}, pages
  177--188, 2016.

\bibitem{Le19}
X.~D. Le, L.~Bao, D.~Lo, X.~Xia, S.~Li, and C.~S. Pasareanu.
\newblock On reliability of patch correctness assessment.
\newblock In {\em ICSE}, 2019.

\bibitem{LeGoues12b}
C.~{Le Goues}, T.~Nguyen, S.~Forrest, and W.~Weimer.
\newblock {GenProg}: {A} generic method for automatic software repair.
\newblock {\em TSE}, 38:54--72, 2012.

\bibitem{LeGoues19}
C.~{Le Goues}, M.~Pradel, and A.~Roychoudhury.
\newblock Automated program repair.
\newblock {\em CACM}, 62(12):56--65, Nov. 2019.

\bibitem{Lee18}
J.~Lee, D.~Kim, T.~F. Bissyand{\'e}, W.~Jung, and Y.~L. Traon.
\newblock {Bench4BL}: {Reproducibility} study on the performance of {IR}-based
  bug localization.
\newblock In {\em ISSTA}, pages 61--72, 2018.

\bibitem{Li19}
X.~Li, W.~Li, Y.~Zhang, and L.~Zhang.
\newblock {DeepFL}: {Integrating} multiple fault diagnosis dimensions for deep
  fault localization.
\newblock In {\em ISSTA}, 2019.

\bibitem{Li17}
X.~Li and L.~Zhang.
\newblock Transforming programs and tests in tandem for fault localization.
\newblock {\em PACML OOPSLA}, 1:1--30, 2017.

\bibitem{Li21}
Y.~Li, S.~Wang, and T.~Nguyen.
\newblock Fault localization with code coverage representation learning.
\newblock In {\em ICSE}, pages 661--673, 2021.

\bibitem{Lima21}
R.~Lima, J.~F. Ferreira, and A.~Mendes.
\newblock Automatic repair of java code with timing side-channel
  vulnerabilities.
\newblock In {\em International Workshop on Refactoring (IWOR)}, pages 1--8,
  2021.

\bibitem{Lin10}
S.~Lin.
\newblock Rank aggregation methods.
\newblock {\em Wiley Interdisciplinary Reviews: Computational Statistics},
  2(5):555--570, 2010.

\bibitem{Liu13}
C.~Liu, J.~Yang, L.~Tan, and M.~Hafiz.
\newblock {R2Fix}: {A}utomatically generating bug fixes from bug reports.
\newblock In {\em ICST}, pages 282--291, 2013.

\bibitem{Liu18icsme}
K.~Liu, D.~Kim, A.~Koyuncu, L.~Li, T.~F. Bissyand{\'{e}}, and Y.~L. Traon.
\newblock A closer look at real-world patches.
\newblock In {\em ICSME}, pages 275--286, 2018.

\bibitem{Liu19}
K.~Liu, A.~Koyuncu, T.~F. Bissyand{\'e}, D.~Kim, J.~Klein, and Y.~L. Traon.
\newblock You cannot fix what you cannot find! an investigation of fault
  localization bias in benchmarking automated program repair systems.
\newblock In {\em ICST}, pages 102--113, 2019.

\bibitem{Liu19issta}
K.~Liu, A.~Koyuncu, D.~Kim, and T.~F. Bissyand\'{e}.
\newblock Tbar: Revisiting template-based automated program repair.
\newblock In {\em ISSTA}, pages 31--42, 2019.

\bibitem{Liu21}
K.~Liu, L.~Li, A.~Koyuncu, D.~Kim, Z.~Liu, J.~Klein, and T.~F. Bissyand{\'e}.
\newblock A critical review on the evaluation of automated program repair
  systems.
\newblock {\em Journal of Systems and Software}, 171:110817, 2021.

\bibitem{Long16}
F.~Long and M.~Rinard.
\newblock Automatic patch generation by learning correct code.
\newblock In {\em POPL}, pages 298--312, 2016.

\bibitem{Lou20}
Y.~Lou, A.~Ghanbari, X.~Li, L.~Zhang, H.~Zhang, D.~Hao, and L.~Zhang.
\newblock Can automated program repair refine fault localization? a unified
  debugging approach.
\newblock In {\em ISSTA}, pages 75--87, 2020.

\bibitem{Mahalakshmi15}
S.~Mahalakshmi and E.~Sivasankar.
\newblock Cross domain sentiment analysis using different machine learning
  techniques.
\newblock In {\em Intl. Conf. on Fuzzy and Neuro Computing (FANCCO)}, pages
  77--87, 2015.

\bibitem{Marginean19}
A.~Marginean, J.~Bader, S.~Chandra, M.~Harman, Y.~Jia, K.~Mao, A.~Mols, and
  A.~Scott.
\newblock {SapFix}: {Automated} end-to-end repair at scale.
\newblock In {\em ICSE}, pages 269--278, 2019.

\bibitem{Martinez17}
M.~Martinez, T.~Durieux, R.~Sommerard, J.~Xuan, and M.~Monperrus.
\newblock Automatic repair of real bugs in {Java}: {A} large-scale experiment
  on the {Defects4J} dataset.
\newblock {\em EMSE}, 22(4):1936--1964, April 2017.

\bibitem{Mechtaev18}
S.~Mechtaev, M.-D. Nguyen, Y.~Noller, L.~Grunske, and A.~Roychoudhury.
\newblock Semantic program repair using a reference implementation.
\newblock In {\em ICSE}, pages 129--139, 2018.

\bibitem{Motwani19icse}
M.~Motwani and Y.~Brun.
\newblock Automatically generating precise oracles from structured natural
  language specifications.
\newblock In {\em ICSE}, May 2019.

\bibitem{MotwaniSBIRArtifact}
M.~Motwani and Y.~Brun.
\newblock Replication data for: {Better} automatic program repair by using bug
  reports and tests together.
\newblock Harvard Dataverse, \url{https://doi.org/10.7910/DVN/OHMYAK}, 2023.

\bibitem{Motwani22}
M.~Motwani, M.~Soto, Y.~Brun, R.~Just, and C.~{Le Goues}.
\newblock Quality of automated program repair on real-world defects.
\newblock {\em TSE}, 48(2):637--661, February 2022.

\bibitem{Noda20}
K.~Noda, Y.~Nemoto, K.~Hotta, H.~Tanida, and S.~Kikuchi.
\newblock Experience report: {How} effective is automated program repair for
  industrial software?
\newblock In {\em SANER}, pages 612--616, 2020.

\bibitem{Noller22}
Y.~Noller, R.~Shariffdeen, X.~Gao, and A.~Roychoudhury.
\newblock Trust enhancement issues in program repair.
\newblock In {\em ICSE}, 2022.

\bibitem{Pearson17}
S.~Pearson, J.~Campos, R.~Just, G.~Fraser, R.~Abreu, M.~D. Ernst, D.~Pang, and
  B.~Keller.
\newblock Evaluating and improving fault localization.
\newblock In {\em ICSE}, pages 609--620, 2017.

\bibitem{Petke20}
J.~Petke and A.~Blot.
\newblock Refining fitness functions in test-based program repair.
\newblock In {\em APR}, page 13–14, 2018.

\bibitem{Pihur09}
V.~Pihur, S.~Datta, and S.~Datta.
\newblock {RankAggreg}, an {R} package for weighted rank aggregation.
\newblock {\em BMC bioinformatics}, 10(1):62, 2009.

\bibitem{Pihur20}
V.~Pihur, S.~Datta, and S.~Datta.
\newblock {RankAggreg}: {Weighted} rank aggregation.
\newblock \url{https://cran.r-project.org/web/packages/RankAggreg/}, 2020.

\bibitem{Qi15}
Z.~Qi, F.~Long, S.~Achour, and M.~Rinard.
\newblock An analysis of patch plausibility and correctness for
  generate-and-validate patch generation systems.
\newblock In {\em ISSTA}, page 24–36, 2015.

\bibitem{ASTExpr}
Rational software architect 9.5.0.
\newblock \url{https://ibm.co/3HfEml7}.
\newblock [accessed 4-March-2022].

\bibitem{Robertson2000}
S.~E. Robertson, S.~Walker, and M.~Beaulieu.
\newblock Experimentation as a way of life: Okapi at trec.
\newblock {\em Information processing \& management}, 36(1):95--108, 2000.

\bibitem{Rubinstein13}
R.~Y. Rubinstein and D.~P. Kroese.
\newblock {\em The {Cross-Entropy Method}: {A} Unified Approach to
  Combinatorial Optimization, Monte-Carlo Simulation and Machine Learning}.
\newblock Springer Science \& Business Media, 2013.

\bibitem{Saha13}
R.~K. Saha, M.~Lease, S.~Khurshid, and D.~E. Perry.
\newblock Improving bug localization using structured information retrieval.
\newblock In {\em ASE}, 2013.

\bibitem{Saha17}
R.~K. Saha, Y.~Lyu, H.~Yoshida, and M.~R. Prasad.
\newblock {ELIXIR}: {Effective} object oriented program repair.
\newblock In {\em ASE}, pages 648--659, 2017.

\bibitem{Sanathanan77}
L.~Sanathanan.
\newblock Estimating the size of a truncated sample.
\newblock {\em Journal of the American Statistical Association},
  72(359):669--672, 1977.

\bibitem{Sanchez-Stern22passport}
A.~Sanchez-Stern, E.~First, T.~Zhou, Z.~Kaufman, Y.~Brun, and T.~Ringer.
\newblock Passport: {Improving} automated formal verification using
  identifiers.
\newblock {\em ACM TOPLAS}, 2023.

\bibitem{Smith15fse}
E.~K. Smith, E.~Barr, C.~{Le Goues}, and Y.~Brun.
\newblock Is the cure worse than the disease? {Overfitting} in automated
  program repair.
\newblock In {\em ESEC/FSE}, pages 532--543, 2015.

\bibitem{Sohn17}
J.~Sohn and S.~Yoo.
\newblock {FLUCCS}: {Using} code and change metrics to improve fault
  localization.
\newblock In {\em ISSTA}, pages 273--283, 2017.

\bibitem{Sun18}
S.~Sun, J.~Guo, R.~Zhao, and Z.~Li.
\newblock Search-based efficient automated program repair using mutation and
  fault localization.
\newblock In {\em COMPSAC}, volume~1, pages 174--183, 2018.

\bibitem{Thomas19science}
P.~S. Thomas, B.~C. {da Silva}, A.~G. Barto, S.~Giguere, Y.~Brun, and
  E.~Brunskill.
\newblock Preventing undesirable behavior of intelligent machines.
\newblock {\em Science}, 366(6468):999--1004, November 2019.

\bibitem{Tian20}
H.~Tian, K.~Liu, A.~K. Kabor\'{e}, A.~Koyuncu, L.~Li, J.~Klein, and T.~F.
  Bissyand\'{e}.
\newblock Evaluating representation learning of code changes for predicting
  patch correctness in program repair.
\newblock In {\em ASE}, 2020.

\bibitem{Tian17}
Y.~Tian and B.~Ray.
\newblock Automatically diagnosing and repairing error handling bugs in {C}.
\newblock In {\em ESEC/FSE}, pages 752--762, 2017.

\bibitem{Tsai07}
M.-F. Tsai, T.-Y. Liu, T.~Qin, H.-H. Chen, and W.-Y. Ma.
\newblock Frank: a ranking method with fidelity loss.
\newblock In {\em SIGIR}, pages 383--390, 2007.

\bibitem{Tufano19}
M.~Tufano, J.~Pantiuchina, C.~Watson, G.~Bavota, and D.~Poshyvanyk.
\newblock On learning meaningful code changes via neural machine translation.
\newblock In {\em ICSE}, page 25–36, 2019.

\bibitem{Wang18}
K.~Wang, R.~Singh, and Z.~Su.
\newblock Search, align, and repair: {Data-driven} feedback generation for
  introductory programming exercises.
\newblock In {\em PLDI}, 2018.

\bibitem{Wang20}
S.~Wang, M.~Wen, B.~Lin, H.~Wu, Y.~Qin, D.~Zou, X.~Mao, and H.~Jin.
\newblock Automated patch correctness assessment: How far are we?
\newblock In {\em ASE}, page 968–980. Association for Computing Machinery,
  2020.

\bibitem{Wen17}
M.~Wen, J.~Chen, R.~Wu, D.~Hao, and S.~Cheung.
\newblock An empirical analysis of the influence of fault space on search-based
  automated program repair.
\newblock {\em CoRR}, abs/1707.05172, 2017.

\bibitem{Wen18}
M.~Wen, J.~Chen, R.~Wu, D.~Hao, and S.-C. Cheung.
\newblock Context-aware patch generation for better automated program repair.
\newblock In {\em ICSE}, 2018.

\bibitem{Wen16}
M.~Wen, R.~Wu, and S.-C. Cheung.
\newblock {Locus}: {Locating} bugs from software changes.
\newblock In {\em ASE}, pages 262--273, 2016.

\bibitem{Wong14}
C.-P. Wong, Y.~Xiong, H.~Zhang, D.~Hao, L.~Zhang, and H.~Mei.
\newblock Boosting bug-report-oriented fault localization with segmentation and
  stack-trace analysis.
\newblock In {\em ICSME}, pages 181--190, 2014.

\bibitem{Wong13}
W.~E. Wong, V.~Debroy, R.~Gao, and Y.~Li.
\newblock The {DStar} method for effective software fault localization.
\newblock {\em TR}, 63(1):290--308, 2013.

\bibitem{Xin17}
Q.~Xin and S.~P. Reiss.
\newblock Identifying test-suite-overfitted patches through test case
  generation.
\newblock In {\em ISSTA}, pages 226--236, 2017.

\bibitem{Xiong18}
Y.~Xiong, X.~Liu, M.~Zeng, L.~Zhang, and G.~Huang.
\newblock Identifying patch correctness in test-based program repair.
\newblock In {\em ICSE}, 2018.

\bibitem{Xiong17}
Y.~Xiong, J.~Wang, R.~Yan, J.~Zhang, S.~Han, G.~Huang, and L.~Zhang.
\newblock Precise condition synthesis for program repair.
\newblock In {\em ICSE}, 2017.

\bibitem{Xuan14icsme}
J.~Xuan and M.~Monperrus.
\newblock Learning to combine multiple ranking metrics for fault localization.
\newblock In {\em ICSME}, pages 191--200, 2014.

\bibitem{Xuan14}
J.~Xuan and M.~Monperrus.
\newblock Test case purification for improving fault localization.
\newblock In {\em FSE}, pages 52--63, 2014.

\bibitem{Yang20}
B.~Yang, Y.~He, H.~Liu, Y.~Chen, and Z.~Jin.
\newblock A lightweight fault localization approach based on xgboost.
\newblock In {\em QRS}, pages 168--179, 2020.

\bibitem{Yang18}
D.~Yang, Y.~Qi, and X.~Mao.
\newblock Evaluating the strategies of statement selection in automated program
  repair.
\newblock In {\em SATE}. Springer, 2018.

\bibitem{Yang17fse}
J.~Yang, A.~Zhikhartsev, Y.~Liu, and L.~Tan.
\newblock Better test cases for better automated program repair.
\newblock In {\em ESEC/FSE}, pages 831--841, 2017.

\bibitem{Ye21}
H.~Ye, M.~Martinez, and M.~Monperrus.
\newblock Automated patch assessment for program repair at scale.
\newblock {\em EMSE}, 26(2), 2021.

\bibitem{Ye22}
H.~Ye, M.~Martinez, and M.~Monperrus.
\newblock Neural program repair with execution-based backpropagation.
\newblock In {\em ICSE}, pages 1506--–1518, 2022.

\bibitem{Youm15}
K.~C. Youm, J.~Ahn, J.~Kim, and E.~Lee.
\newblock Bug localization based on code change histories and bug reports.
\newblock In {\em APSEC}, 2015.

\bibitem{Yu19}
Z.~Yu, M.~Martinez, B.~Danglot, T.~Durieux, and M.~Monperrus.
\newblock Alleviating patch overfitting with automatic test generation: {A}
  study of feasibility and effectiveness for the {Nopol} repair system.
\newblock {\em EMSE}, 24(1):33--67, 2019.

\bibitem{Yuan20}
Y.~Yuan and W.~Banzhaf.
\newblock {ARJA}: {Automated} repair of {Java} programs via multi-objective
  genetic programming.
\newblock {\em TSE}, 46(10):1040--1067, 2020.

\bibitem{Zhang17}
M.~Zhang, X.~Li, L.~Zhang, and S.~Khurshid.
\newblock Boosting spectrum-based fault localization using pagerank.
\newblock In {\em ISSTA}, pages 261--272, 2017.

\bibitem{Zhou12}
J.~Zhou, H.~Zhang, and D.~Lo.
\newblock Where should the bugs be fixed? more accurate information
  retrieval-based bug localization based on bug reports.
\newblock In {\em ICSE}, pages 14--24, 2012.

\bibitem{Zhu21}
Q.~Zhu, Z.~Sun, Y.~an~Xiao, W.~Zhang, K.~Yuan, Y.~Xiong, and L.~Zhang.
\newblock A syntax-guided edit decoder for neural program repair.
\newblock In {\em ESEC/FSE}, pages 341--353, 2021.

\bibitem{Zou19}
D.~Zou, J.~Liang, Y.~Xiong, M.~D. Ernst, and L.~Zhang.
\newblock An empirical study of fault localization families and their
  combinations.
\newblock {\em TSE}, 2019.

\bibitem{combineFL}
D.~Zou, J.~Liang, Y.~Xiong, M.~D. Ernst, and L.~Zhang.
\newblock Evaluating and combining fault localization techniques from different
  families.
\newblock \url{https://damingz.github.io/combinefl}, 2019.
\newblock [accessed 4-March-2022].

\end{thebibliography}

\end{document}